%% file: 3pc_r+k_space.tex
\documentclass[useAMS,twocolumn,usenatbib]{mnras}
\usepackage{amssymb}
\usepackage{amsmath}
\usepackage{grffile}
\usepackage{graphicx}
\usepackage{caption}
\usepackage[british]{babel} 
\usepackage{placeins}
\usepackage{color}
\usepackage{natbib}
\usepackage{appendix}
\usepackage[T1]{fontenc}
\usepackage{ae,aecompl}

\newcommand{\mpc}[1]{$h^{-1}\text{Mpc}$}
\newcommand{\rtr}[1]{$(r_1r_2r_3)^{1/3}$}
\newcommand{\degree}{\ensuremath{^\circ}}
\newcommand{\bdsqfix}{$b_{\delta^2 fix}$}
\newcommand{\css}{CSS12}

\title[Halo 3PCF in Fourier and configuration space]{Testing the consistency of three-point halo clustering in Fourier and configuration space}

\author[Hoffmann et al.]{
K. Hoffmann,$^{1,2}$ \thanks{E-mail: hoffmann@tsinghua.edu.cn}
E. Gazta\~naga$^{2},$
R. Scoccimarro$^{3},$
M. Crocce$^{2}$
\\
$^{1}$Tsinghua Center for Astrophysics, Department of Physics, Tsinghua University, Beijing, 100084, China\\
$^{2}$Institute  of  Space  Sciences,  IEEC-CSIC, Campus  UAB, Carrer  de  Can  Magrans,  s/n,  08193  Barcelona,  Spain \\
$^{3}$Center for Cosmology and Particle Physics, Department of Physics, New York University, NY 10003, New York, USA
}

\date{Accepted XXX. Received YYY; in original form ZZZ}
\pubyear{2017}

\begin{document}
\label{firstpage}
\pagerange{\pageref{firstpage}--\pageref{lastpage}}
\maketitle

\begin{abstract}
We compare reduced three-point correlations $Q$ of matter, haloes (as proxies for galaxies) and their cross correlations,
measured in a total simulated volume of $\sim100 \ (h^{-1} \text{Gpc})^{3}$, to predictions from leading order perturbation theory
on a large range of scales in configuration space.
Predictions for haloes are based on the non-local bias model, employing linear ($b_1$) and non-linear ($c_2$, $g_2$)
bias parameters, which have been constrained previously from the bispectrum in Fourier space.
We also study predictions from two other bias models, one local ($g_2=0$) and one in which $c_2$ and $g_2$ are determined by $b_1$
via approximately universal relations. Overall, measurements and predictions agree when $Q$ is derived for triangles
with $(r_1r_2r_3)^{1/3} \gtrsim 60 h^{-1}\text{Mpc}$, where $r_{1-3}$ are the sizes of the triangle legs.
Predictions for $Q_{matter}$, based on the linear power spectrum, show significant deviations from the measurements
at the BAO scale (given our small measurement errors), which strongly decrease when adding a damping term or using the non-linear
power spectrum, as expected. Predictions for $Q_{halo}$ agree best with measurements at large scales when considering non-local contributions.
The universal bias model works well for haloes and might therefore be also useful for tightening constraints on $b_1$ from $Q$
in galaxy surveys. Such constraints are independent of the amplitude of matter density fluctuation ($\sigma_8$) and hence
break the degeneracy between $b_1$ and $\sigma_8$, present in galaxy two-point correlations.
\end{abstract}
 

\input{sections/introduction.tex}

\input{sections/3pc_sim-pt.tex}
\input{sections/accuracy_of_Qpredictions.tex}

\input{sections/conclusions.tex}


\FloatBarrier
\section*{Acknowledgements}
We acknowledge support from the Spanish Ministerio de Ciencia e Innovacion (MICINN) projects AYA2012-39559 and AYA2015-71825,
and research project 2014 SGR 1378 from the Generalitat de Catalunya.
KH acknowledges the support by the International Postdoc Fellowship from the Chinese Ministry of Education and the State Administration of Foreign Experts Affairs.
He also thanks the organisers and participants of the 2016 workshop {\it Biased Tracers of Large-Scale Structure} at the Lorentz Center as well as
Kwan Chuen Chan for useful discussions. MC acknowledges  support from the MICINN project AYA2013-44327 and the Ramon y Cajal program.

\bibliographystyle{mnras}
\bibliography{references.bib}

\appendix
\input{./sections/appendix.tex}


\bsp	
\label{lastpage}
\end{document}

%% file: sections/introduction.tex
\section{Introduction}
Higher-order correlations, induced by gravity into the distribution of large-scale matter density fluctuations,
contain information which cannot be captured by second-order statistics.
This information can be used to tighten constraints on cosmological models, as well as on
models of galaxy formation.
A key tool for obtaining such constraints are galaxy bias models \citep[e.g.][]{DJS16}.
These models relate the density and the tidal field of the full matter content in a given region
to the density of observable tracers, such as galaxies.
They include a number of so-called bias parameters, which depend on the various processes that drive the tracer formation.
Since these highly complex processes are only partly understood, the bias parameters cannot be predicted in a reliable way 
\citep[e.g.][]{Li07, mueller11, Pujol17, Springel17} and hence need to be measured from the data.
Such measurements can be obtained from the analysis of weak gravitational lensing signals, or redshift space
distortions. However, these methods rely on good redshift estimations and imaging of the tracers (i.e. galaxies) as well as on
various model assumptions. 
It is therefore interesting to obtain independent measurements of the bias parameters, which is possible with a joint
analysis of second- and third-order statistics.
This approach becomes increasingly interesting as errors on these statistics decrease with the increasing volumes of upcoming galaxy surveys.
Going to third-order in the statistical analysis of galaxy surveys does not only deliver bias measurements,
but also measurements of the growth of matter fluctuations. The latter provide the aforementioned
cosmological constraints, while the bias can be used to predict the number of galaxies per halo, which
places constraints on galaxy formation models \citep[e.g.][]{Scoccimarro01, Berlind2002, CooraySheth02}.

The most general third-order statistics
is the three-point correlation function (hereafter referred to as 3PCF), which is defined in configurations space.
Alternatively one can study its Fourier space counterpart, the bispectrum.
These two statistics contain in principle the same information. However, their analyses implicate
different limitations and challenges, which can affect the physical interpretation of the results.
A main advantage of the bispectrum is that an analysis in Fourier space allows for a clear exclusion of high 
frequency modes in the density fluctuations, which are difficult to interpret theoretically due to their highly
non-linear evolution. In configuration space, these high frequency modes contribute to the 3PCF
in principle at all scales. In practice one therefore needs to restrict the analysis to large scales, where
their contribution is negligible, lavishing a lot of valuable data.
Another advantage of the bispectrum is that its covariance is diagonal for Gaussian density fluctuations.
This approximation works well, even for evolved density fields, while deviations from Gaussianity can also
be taken into account \citep{Scoccimarro00, Sefusatti06, ChaBlo16}. The covariance of the 3PCF on the other hand is not diagonal, even
for Gaussian fluctuations, which makes the modeling more difficult \citep{Sred93, SE15, Byun17, Gualdi17}.
An additional difference in the analysis of the bispectrum and the 3PCF lies in the fact that the computation of the
latter is more expensive. However, this aspect can be tackled by employing advanced algorithms and appropriate
computational resources, as done in this work \citep[see also, ][and references therein]{bargaz, mcbride11a, Jarvis15, SE15}.
Besides its disadvantages, there are some arguments which speak for the 3PCF. One of them is the
fact that the amplitude of the 3PCF (but not its errors) is not affected by shot-noise, whereas the latter affects
the bispectrum amplitude at all scales and hence needs to be modelled for correcting the measurements.
In addition, an analysis in configuration space has the advantage that complicated survey masks
can be easily taken into account in the analysis of observational data, while in Fourier space such masks
impose complicated effects on the measured bispectrum, which are difficult to model \citep[e.g.][]{Scoccimarro00}.
A more general consideration is that it is easier to interpret effects such as redshift space distortions or
baryon acoustic oscillations (BAO) on the statistics in configuration space, since that is where the physical
processes which cause these effects happen.
Studies of third-order correlations in the literature usually focus on either Fourier or configuration space
\citep[e.g.][]{mcbride11b, marin13, GilMarin15}.
However, it is worthwhile studying both statistics and cross-check the results, since their different
advantages and disadvantages are quite complementary.

In this work we will conduct such a cross-check for the first time. Our main interest thereby is to verify if
and when the bias parameters, obtained from the bispectrum are consistent with those which affect the 3PCF in configuration space.
Our approach is based on the analysis of \citet[][hereafter referred to as \css]{chan12}. These authors measured the bias parameters
of large-scale structure tracers in Fourier space from a set of N-body simulations, using a leading-order perturbative model of the bispectrum
and restricting the analysis to large modes with wave numbers $k \le 0.1 \ h \text{Mpc}^{-1}$.
The tracers in their analysis are dark-matter halos, while the same method for measuring the bias can be applied to
any other type of tracers, such as galaxies or galaxy clusters.
For our cross-check we use the same perturbative model together with the bias parameters of \css{}
to predict the halo 3PCF  in configuration space. We then measure the latter in the same set of simulations to test the predictions.
This allows us to verify if and when the bias parameters measured from third-order statistics in Fourier space also describe the
corresponding statistics in configuration space.
Simultaneously we test at which scales, redshifts and halo mass ranges the leading order perturbative modeling of the 3PCF is an
appropriate approximation.

\subsection{Bias models tested}
\label{sec:bias_models}

The bias model relates the density fluctuations and the tidal field of matter in a certain region to the density fluctuations of its tracers.
These fluctuations are defined with respect to the mean density as $\delta \equiv (\rho - \bar{\rho})/\bar{\rho}$.
Since the leading order perturbative expansion of third-order statistics, on which we focus in this analysis, is quadratic,
we use the quadratic non-local bias model,
\begin{equation}
	\delta_h= b_1  \biggl\{ \delta_m + \frac{1}{2}[ c_2(\delta_m^2 - \langle \delta_m^2  \rangle) + g_2  \mathcal{G}_2] \biggr\}.
	\label{eq:bias_model}
\end{equation}
The indices $h$ and $m$ refer to the halo and matter density fluctuations respectively. 
The parameters $b_1$ and $c_2$ are hereafter referred to as local linear and quadratic bias \citep{FG}, while $g_2$ will be
referred to as quadratic non-local bias, since it scales with the tidal field term $\mathcal{G}_2$,
which can be generated by masses outside of the volume in which $\delta_g$ is defined \citep[see \css{};][]{McDonald09, baldauf12}.
The term for the smoothed tidal field is given by a second-order Gallileon
\begin{equation}
  \mathcal{ G}_2({\bf r})= \int \beta_{12}\theta_v({\bf k}_1) \theta_v({\bf k}_2) \
  \hat W[k_{12}R]e^{i {\bf k}_{12}\cdot {\bf r} }d^3 {\bf k}_1 d^3 {\bf k}_2,
\label{eq:G2}
\end{equation}
where ${\bf k}_i$ and ${\bf k}_{12} \equiv {\bf k}_2 - {\bf k}_1$ are wave vectors of density oscillations,
$\beta_{12} \equiv 1 - ({\hat{\bf k}}_1 \cdot {\hat{\bf k}}_2)^2$ represents the mode-coupling between
density oscillations which describes tidal forces, $\theta_v \equiv \nabla^2 \Phi_v$ is the divergence of the
normalised velocity field (${\bf v}/\mathcal{H}/f$) and  $ \hat W[k_{12}R]$ is the window function in Fourier space (\css).
Note that the non-local bias has also been referred to as an additional local bias
parameter, since the tidal field is a local observable, which depends on derivatives of the potential \cite[see \css{},][]{DJS16}.
However, in this work we call it non-local, since it is non-local in the density.

We use three sets of bias parametrizations for predicting the
3PCF. The {\it first set} consists of the bias parameters $b_1$, $c_2\equiv b_2/b_1$
and $g_2\equiv 2\gamma_2/b_1$, obtained by \css{} from fitting the non-local bias model predictions for the bispectrum
at leading order to measurements in the same set of simulations as studied in this work. Here $b_2$ and $\gamma_2$
are the quadratic local and the non-local bias parameters respectively, in the notation of \css{}.
The {\it second set} equals the first set, except for the non-local bias parameter $g_2$, which is set to zero in order to verify the impact
of the non-local contributions on the 3PCF predictions. In the {\it third set} the linear bias $b_1$ is the only input parameter. This
is the same parameter, as in the two previous sets, but was obtained by \css{} from fits of the linear bias model ($\delta_h = b_1 \delta_m$) to the power spectrum.
The quadratic local bias parameter $c_2$ in this third set  is computed from the (approximately) universal relation $c_2 \simeq 0.77 b_1^{-1} - 2.43 + b_1$ , given by
\citet{HBG17} \citep[see also][for similar relations]{HBG15b,  Lazeyras15}. The non-local bias in the third set is obtained from the local Lagrangian model, $g_2 = -(4/7)(1-1/b_1)$.
%
The three sets of bias model parameters are summarised in Table \ref{table:bias_models} and will in the following
be referred to as {\it non-local}, {\it local} and \bdsqfix{} model respectively.
\begin{table}
\centering
\caption{Different bias models studied in this work. The bias parameters have been measured in Fourier space from
the same set of simulations by \css. For the \bdsqfix{} model we use a roughly universal relation for the quadratic local bias
$c_2(b_1)$ \citep{HBG17} and the local Lagrangian model for the quadratic non-local $g_2(b_1)$.}
\label{table:bias_models}
\begin{tabular}{cccc}
bias model & description \\\hline
non-local &  $b_1$, $c_2$, $g_2$ from bispectrum fits\\
local &  same $b_1$ and $c_2$ as above, $g_2=0$   \\
\bdsqfix{} & $b_1$ from power spectrum fits, \\
&$c_2 = 0.77 b_1^{-1} - 2.43 + b_1$, \\
&$g_2 = -(4/7)(1-1/b_1)$
\end{tabular}
\end{table}

For a three-dimensional analysis of real spectroscopic
surveys one would further need to take into account redshift space distortions in the modeling.
Redshift space distortions cancel out approximately at large scales in the reduced 3PCF
\citep[defined in Section \ref{subsec:3pc_def}, see e.g.][] {GazSco05}, but there are non-linear contributions
that could be as large as the non-local terms. There is some
indication in simulations that non-local terms can cancel out with
redshift space distortions, \citep[e.g. Fig.17 in][]{HBG15a}, but this
requires further study.
Large volume photometric surveys, such as DES or LSST will provide additional constraints from weak lensing of the projected 3PCF, both of galaxy and matter 
correlations, as well as galaxy-matter cross-correlations. Since those surveys measure redshifts from broad-band photometry these probes will have little contamination 
by redshift space distortions. All this is beyond the scope of this paper, but should be a clear continuation of our study.

%% file: sections/3pc_sim-pt.tex
\section{Three-point correlations $Q$}\label{sec:3pc}

\subsection{Definitions}\label{subsec:3pc_def}

Our 3PCF analysis is applied on density fields $\rho^x(\bf r)$, where $x$ refers to the density of matter ($x=m$) or of its tracers, such as galaxies
or, as in our case, dark matter halos ($x=h$) at the position $\bf r$. The density fields are smoothed with a top-hat filter of scale $R$ and described
by the normalised density fluctuations $\delta^{x}({\bf r}_i) \equiv \delta^{x}_i$, introduced in Section \ref{sec:bias_models}. Note that,
in contrast to the notation in equation (\ref{eq:bias_model}), we now set $x$ as upper index to avoid confusion between the position and the power
indices in the following. The 3PCF can be defined as the average product of density fluctuations at three positions
$({\bf r}_{1}, {\bf r}_{2}, {\bf r}_{3})$, which form a triangle. In the case of the halo-matter-matter cross-correlations it is written as
\begin{equation}
		\zeta^{hmm} (r_{12}, r_{13}, r_{23}) \equiv \langle  \delta^h_1 \delta^m_2 \delta^m_3 \rangle(r_{12}, r_{13}, r_{23}),
		\label{eq:3pcc_def}
\end{equation}
where $r_{ij} \equiv |{\bf r}_i - {\bf r}_j|$ are the absolute values of the triangle legs and $\langle \ldots \rangle$
denotes the average over all possible triangle orientations and translations. We proceed by defining the
symmetric reduced three-point cross-correlation, 
\begin{equation}
	Q_{\times} \equiv \\
	\frac{1}{3}
	\frac
	{\zeta^{hmm}  + \zeta^{mhm} + \zeta^{mmh}}{\zeta^{\times} _H}
	\label{eq:Qx_def}
\end{equation}
and drop the expression {\it reduced} in the following. The hierarchical three-point cross-correlation in the denominator, defined as
\begin{equation}
	\zeta^{\times} _H \equiv \xi^{hm}_{12} \xi^{hm}_{13}  + \xi^{hm}_{12} \xi^{hm}_{23}  + \xi^{hm}_{13} \xi^{hm}_{23},
	\label{eq:def_3pcH}
\end{equation}
is comprised of two-point cross-correlations $\xi_{ij}^{xy} \equiv \langle \delta^x_i \delta^y_j \rangle(r_{ij})$ between the density fields $x$ and $y$
\citep[e.g.][]{Peebles75, Fry84}. The corresponding expressions for the three-point auto correlations for matter and halos, ($Q_m$ and $Q_h$ respectively) are defined
analogously, i.e. $Q_m \equiv \zeta^{mmm} / \zeta_H^{mmm}$ and $Q_h \equiv \zeta^{hhh} / \zeta_H^{hhh}$.

\subsection{Modeling}\label{subsec:3pc_model}

Our predictions for $Q_h$ and $Q_\times$ are based on the non-local quadratic bias model from equation (\ref{eq:bias_model}), which yields at
leading-order perturbative expansion in terms of $\delta_m$
\begin{equation}
	Q_h \simeq \frac{1}{b_1}[Q_m + c_2 +  g_2 Q_{nloc}],
	\label{eq:Qh_model}
\end{equation}
where $b_1$ and $c_2$ are the local linear and quadratic bias parameters respectively. The non-local
contribution  $Q_{nloc}$ scales with the non-local quadratic bias parameter $g_2$
\citep[see \css{};][]{baldauf12}. The corresponding leading order expression
for the halo-matter-matter cross-correlation is given by
\begin{equation}
	Q_\times \simeq \frac{1}{b_1}[Q_m + \frac{1}{3}(c_2 +  g_2 Q_{nloc})].
	\label{eq:Qx_model}
\end{equation}
Equation (\ref{eq:Qh_model}) has an important application in the analysis of galaxy surveys,
since it allows for bias measurements which are independent of the linear growth of matter
fluctuations \citep[e.g.][]{friga1994, Sefusatti06, mcbride11b, marin13, GilMarin15} and hence breaks
the growth-bias degeneracy. However, cosmological constraints from such bias measurements are limited
by the inaccuracies of the $Q_h$ modeling as explained in the following.

The statistics of the full matter field, $Q_m$ and $Q_{nloc}$, cannot be observed
in galaxy surveys and hence need to be predicted for a given cosmology. $Q_m$ is therefore often predicted from N-body simulations.
This approach has also been used by \css{} for measuring the bias parameters in their simulations, as it captures the non-linear
contributions to $Q_m$. However, these authors employ an analytical expression for the quadratic non-local contribution
$Q_{nloc}$, which is in Fourier space simply related to the cosine of the angle between two wave vectors.
Direct measurements of $Q_{nloc}$ would be more complicated (see Section \ref{subsec:accuracy_dQ}).
Another disadvantage of deriving the $Q_m$ and $Q_{nloc}$ from simulations, is that a dense sampling of the cosmological parameter
space for deriving constraints from observations would require enormous resources
(albeit $Q_m$ and $Q_{nloc}$ are independent of the linear growth factor and hence only weakly depend on cosmology at large scales).
In this analysis we will therefore employ predictions from leading order perturbation theory for $Q_m$ and $Q_{nloc}$
\citep{Jing97, Gaztanaga98, bargaz, BHG15}. These leading order approximations, as well as those of $Q_h$ and $Q_\times$
in equation (\ref{eq:Qh_model}) and (\ref{eq:Qx_model}) introduce inaccuracies in the modeling, in particular at small triangle scales,
which are strongly affected by high frequency, non-linear modes \citep{Scoccimarro98, Pollack12}.

Measurements of the different 3PCFs in N-body simulations allow us to validate these approximations.
For comparing $Q_h$ and $Q_\times$ with such measurements, we employ bias parameters measured from the power spectrum and the bispectrum in Fourier space by \css{} in the
same set of simulations as used in this analysis. We thereby do not only test the validity of the perturbation theory predictions for the 3PCFs,
but also if the bias parameters in Fourier and configuration space are consistent with each-other. Note that the Fourier space bias measurements are also
based on leading order perturbation theory predictions for the cross-bispectrum $B_{hmm}$ and the non-local contribution. However, non-linear contributions can be
excluded in that case in a more reliable way than in configuration space by restricting the analysis to long wavelength modes. We therefore consider them to be robust.

To summarise, the accuracy of the model of $Q_h$ in equation (\ref{eq:Qh_model}) depends on the accuracy of the leading order
perturbative expansion of $Q_h$, $Q_m$ and $Q_{nloc}$. The comparison with measurements in simulations will further depend on
the accuracy of the bias measurements in Fourier space. In Section \ref{sec:accuracy_Q}
we will test these different model ingredients using measurements of $Q_m$, $Q_h$ and $Q_\times$.

\subsection{Measurements in simulations}\label{subsec:3pc_sim}

We verify the model predictions using the same set of $N_{sim}=49$ cosmological N-body simulations, which was analysed by \css{}.
Each simulation was run with $640^3$ dark matter particles, which reside in a cube with comoving side length of $1280$ \mpc{},
which results in a total simulated volume of $\sim100 \ (h^{-1} \text{Gpc})^{3}$.
The cosmological parameters were set to $\Omega_m= 1- \Omega_{\Lambda} = 0.27$, with $\Omega_b = 0.046, h=0.72, n_s=1$
and $\sigma_8=0.9$. Halos were identified as {\it friends-of-friends} groups with a linking length of $0.2$ of the mean
particle separation. We split them into the same mass samples as \css{}, which are summarised in Table \ref{table:nyu_halo_masses}.

For measuring the 3PCFs in these simulations we generate density maps of the simulated halo and matter distributions
based on $8 \ h^{-1}$Mpc cubical cells. The products of density contrasts $\delta_1 \delta_2 \delta_3$,
over which we average to compute $Q$, are obtained from triplets of these cells, which we
find using an algorithm described by \citet{bargaz}. This algorithm delivers measurements for triangle configurations,
defined by the fixed leg sizes $r_{1}, r_{2}$ at different opening angles
$\alpha \equiv \arccos( \hat{\bf r}_1 \cdot \hat{\bf r}_2)$.
The fixed triangle legs are defined with a tolerance $r_i\pm \delta r$, while we set $\delta r$ to values between $1$ and $4$ \mpc{},
depending on the triangle configuration. This tolerance is needed for finding a large number of triplets on the grid
and thereby reduce the impact of shot-noise on the 3PCF measurements. We study the impact of this tolerance
on the 3PCF, by computing the 3PCF predictions for the same set of triangles, which we find on the grid
for a given $(r_1, r_2) \pm (\delta r_1, \delta r_2)$. We then bin the results for different opening angles $\alpha$ and
compare them to predictions for exact $(r_1, r_2)$ values, as we use them in our analysis (see Appendix \ref{app:3PCF_binning}).
This comparison shows, that the effect of this tolerance and the binning on the 3PCF are small, compared to inaccuracies of
the 3PCF predictions.
The 3PCFs are computed for $28$ configurations ($r_{1}, r_{2}$), with $18$ opening angles each, which leads to a total
number of $504$ triangles.

\begin{table}
\centering
  \caption{Halo mass samples with corresponding linear bias from the halo-matter cross-power spectrum
  from \css{}. The same samples are used in this work.}
  \label{table:nyu_halo_masses}
  \begin{tabular}{c c c c c}
 \hline
 $z$& halo sample  & mass range [$10^{13} M_{\odot}/h$] & $b^P_{hm}$\\
     \hline 
$0.0$	&	m0		&	$4 - 7$ 		&	$1.43$  \\
$0.0$	&	m1		&	$7- 15$ 		&	$1.75$  \\
$0.0$	&	m2		&	$> 15$ 		&	$2.66$ \\
      \hline
$0.5$	&	m0		&	$3 - 5$ 		&	$1.88$ \\
$0.5$	&	m1		&	$5 - 10$ 		&	$2.26$ \\
$0.5$	&	m2		&	$> 10$		&	$3.29$\\
      \hline
$1.0$	&	m0		&	$2 - 3.1$ 		&	$2.43$ \\
$1.0$	&	m1		&	$3.1- 5.7$ 	&	$2.86$ \\
$1.0$	&	m2		&	$> 5.7$		&	$3.99$ \\
      \hline
   \end{tabular}
\end{table}

\subsection{Error estimation}\label{subsec:3pc_errorrs}

To quantify the deviations between the mean 3PCF measurements from the $49$ simulations, $\bar Q_i$,
and the corresponding model predictions, $Q_i^{mod}$, for a set of $N_{\nabla}$ triangles
(each defined by $r_1$, $r_2$ and $\alpha$, with $i \in \{1,2,\ldots,N_{\nabla}\}$), we want to compute
\begin{equation}
\chi^2=\sum^{N_{\nabla}}_{ij} \Delta_i {\hat C}_{ij}^{-1}\Delta_j,
\label{eq:chisq}
\end{equation}
where $\Delta_i \equiv (Q^{mod}_i-\bar Q_i)/\sigma_i$. The standard deviation
of $\bar Q_i$ is given by $\sigma_i^2 = \langle (Q_i - \bar Q_i)^2 \rangle / N_{sim}$, while
$\langle \ldots \rangle$ denotes the mean over the $N_{sim}$ measurements.
The factor $1/N_{sim}$  accounts for the fact that we study the deviations of the mean
measurements from the prediction, rather than deviations of measurements in individual realisations. 
The normalised covariance (or correlation) matrix is hence given by
\begin{equation}
	\hat C_{ij} = \langle \Delta_{i}\Delta_{j} \rangle /N_{sim},
	\label{eq:cov}
\end{equation}
with $\Delta_{i} \equiv (Q_i - \bar Q_i)/\sigma_i$.
We choose $N_{\nabla} < N_{sim}$ to allow for the inversion of $C_{ij}$, as pointed out by
\citet{Hartlap07}, and set $N_{\nabla}=30$ for the $\chi^2$ measurements shown in this paper.
We tested that these measurements are consistent, but
noisier (less noisy) when setting  $N_{\nabla} = 20 \ (40)$, which presumably
results from the relatively low number of $49$ realisations.
To reduce this noise, we follow \citet{GazSco05} by performing a Singular
Value Decomposition of the covariance (hereafter referred to as SVD), i.e.
\begin{equation}
	\hat C_{ij} = (U_{ik})^{\dagger} D_{kl} V_{lj}.
	\label{eq:covSDV}
\end{equation}
The diagonal matrix $D_{kl} = \delta_{kl}\lambda_k^2$ consists of 
the singular values $\lambda_j$ (SVs), while the corresponding normalised
modes ${\bf \hat M}_i$ form the matrix $U$. The modes associated to the largest SVs
may be understood analogously to eigenvectors. We tested that they
build a nearly orthogonal basis, in which we can approximate equation
(\ref{eq:chisq}) as
\begin{equation}
	\chi^2 \simeq \sum_i^{N_{mode}} \langle \Delta_{i} {\bf \hat M}_i  \rangle^2 / \lambda_i^2.
	\label{eq:chisq_SVD}
\end{equation}
Note that here $\langle \ldots \rangle$ denotes the scalar product, i.e. the projection
of $\Delta_{i}$ on ${\bf \hat M}_i$, while $\Delta_{i}$ is the same quantity which
appears in equation (\ref{eq:chisq}).
Fig. \ref{fig:covSV} shows that $\hat C_{ij}$ is typically dominated only by a few modes.
Assuming that the modes with the lowest SVs can be associated with measurement noise,
we use only SVs with values larger than the sampling error estimate
 (i.e. $\lambda^2 \gtrsim \sqrt{2/N_{sim}}$) for our $\chi^2$ computation,
as suggested by \citet{GazSco05}. The number of selected modes is hence the
degree of freedom in our $\chi^2$ estimation, i.e. d.o.f. = $N_{mode} < N_{bin} < N_{sim}$.

%% file: sections/accuracy_of_Qpredictions.tex
\section{accuracy of $Q$ predictions}\label{sec:accuracy_Q}

The accuracy of bias measurements from the reduced three-point halo auto-correlation $Q_h$ in observations depends on
how well it is approximated by the leading order perturbative model, given by equation (\ref{eq:Qh_model}).
To verify this model we test its different components separately with direct measurements in the simulations described in Section \ref{subsec:3pc_sim}.
We start with testing the modeling of $Q_m$ in Section \ref{subsec:accuracy_Qm} and proceed in Section \ref{subsec:accuracy_dQ}
with tests of the quadratic component in equation (\ref{eq:Qh_model}). Our measurements of the latter are obtained by
combining the three-point auto- and cross-correlations, $Q_h$ and $Q_\times$ respectively. Finally, we compare the complete predictions for
$Q_h$ and $Q_\times$, given by equation (\ref{eq:Qh_model}) and (\ref{eq:Qx_model}), with the measurements in the simulations
in Section \ref{subsec:accuracy_QhQx}. For modeling the quadratic components we use bias parameters measured by \css{} in
the same set of simulations, using a leading order perturbative approximation of the 3PCF in Fourier space,
i.e. the tree-level bispectrum. In addition we employ simple relations between the linear and the quadratic bias parameters,
i.e. $b_2(b_1)$ and $g_2(b_1)$. This leaves $b_1$ as the only free input parameter in the bias model, which we
adopt from the fits to the power spectrum, given by \css{} (see Section \ref{sec:bias_models}).

\begin{figure}
  \centering
   \includegraphics[width=70mm, angle = 270]{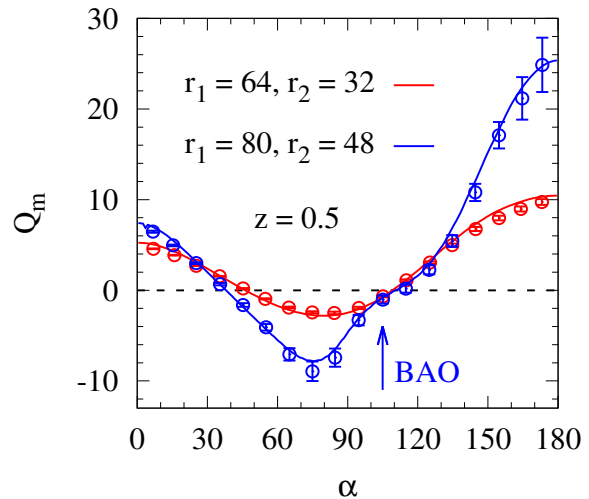}
   \caption{Reduced matter 3PCF $Q_m$ for triangles with fixed legs $r_1$ and $r_2$
   (size is indicated in \mpc{}) versus the mean triangle opening angle $\alpha \equiv \arccos( \hat{\bf r}_1 \cdot \hat{\bf r}_2)$ in each bin. Symbols show mean
   measurements from $49$ simulations with $1\sigma$ errors at redshift $z=0.5$. Lines show tree-level predictions
   from the measured (non-linear) power spectrum.}
\label{fig:Qm_alpha}
\end{figure}

\begin{figure}
  \centering
   \includegraphics[width=120mm, angle = 270]{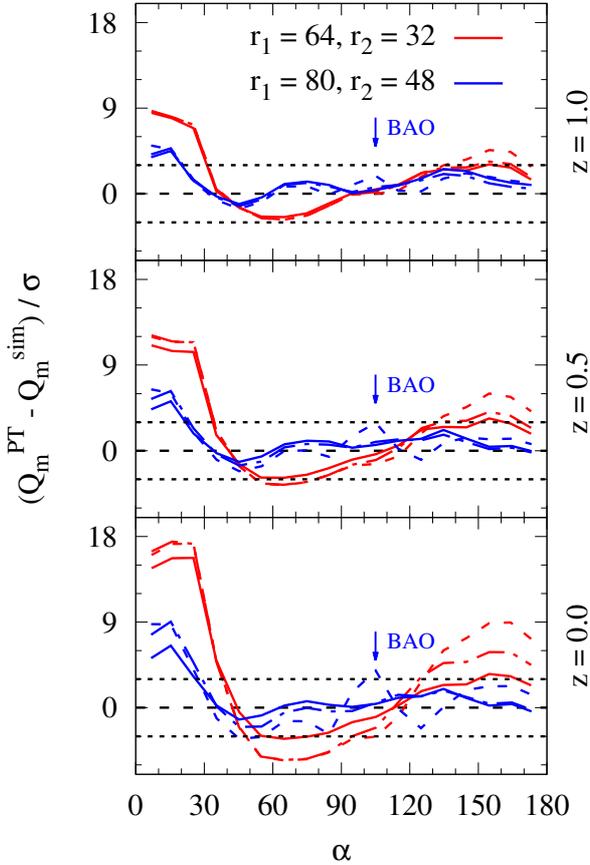}
   \caption{Significance of the deviations between the mean reduced matter 3PCF, measured in $49$ simulations
   and different tree-level predictions for the same triangle configurations as shown in Fig. \ref{fig:Qm_alpha}.
   Dashed and solid lines show predictions from the linear and the measured (non-linear) power spectrum respectively.
   Predictions based on the de-wiggled power spectrum are shown as dashed-dotted lines. The black dotted lines
   mark $3\sigma$ deviations from the measurements.}
\label{fig:Qm_signif_alpha}
\end{figure}

\subsection{$Q_m$}\label{subsec:accuracy_Qm}

We start our verification of the  $Q_m$ model from leading order (tree level) perturbation theory
(hereafter referred to as $Q_m^{PT}$, see Section \ref{subsec:3pc_model})
by comparing its predictions to measurements in simulations. As examples we show in Fig. \ref{fig:Qm_alpha} results at redshift $z=0.5$
for triangles with fixed legs $(r_1,r_2)=(64,32)$ and $(80,48)$ \mpc{} versus the triangle opening angle $\alpha \equiv \arccos( \hat{\bf r}_1 \cdot \hat{\bf r}_2)$.
The $Q_m$ measurements are the mean results of the $49$ simulations and are shown with $1 \sigma$ errors bars.
Here and throughout the paper we display measurements at the mean opening angle in each bin.
The predictions in Fig. \ref{fig:Qm_alpha} are computed from the non-linear power spectrum, which was measured in the simulations.
Both, measurements and predictions exhibit a u-shape, which is more strongly pronounced for the larger triangle configuration and originates
from the filamentary structure of the cosmic web. The measurements clearly show the baryon acoustic oscillations (BAO) feature for the $(80,48)$ \mpc{}
configuration at around $105\degree$.  Indications for similar BAO 3PCF features in real data have first been reported for luminous red galaxies 
in the SDSS DR7 sample by \citet{Gaztanaga09}. \citet{Slepian15} later reported indications for the 3PCF BAO feature in the SDSS DR12 BOSS CMASS sample, 
which were comfirmed by the $4.5\sigma$ detection in the same data set by \citet{Slepian17}.

The significance of the deviations between measurements and predictions is shown for all redshifts in Fig. \ref{fig:Qm_signif_alpha}.
In addition to the predictions from the non-linear power spectrum, we show in this figure also results based on the linear as well as
the so-called de-wiggled power spectrum (hereafter also referred to as $P_{lin}$ and $P_{dw}$ respectively). The latter introduces
non-linearities around the BAO scale in the 3PCF, coming from large-scale displacements \citep{CroSco08, Carlson13, Baldauf2015b, SeZa15, Blas16}.
It consists of the no-wiggle approximation of the power spectrum ($P_{nw}$) from \citet{EH98}, $P_{lin}$ and a smearing function,
i.e., $P_{dw} \equiv P_{nw} + (P_{lin}-P_{nw})exp(-k^2 \sigma_v)$, where $k$ is the wave number and $\sigma_v$ is the variance of the
displacement field \citep[][]{Eisenstein07}\footnote{
We find that replacing the velocity dispersion with the quantity $\Sigma^2(r_{BAO}) \equiv \int_0^\Lambda P(k)d^3k/(3k^2)[1-j_0(k r_{BAO})+2j_2(k r_{BAO})]$
\citep[with $j_n$ being the spherical Bessel function of order $n$,][]{Baldauf2015b}, has a negligibly small effect on the de-wiggled predictions, compared to the
deviations from other predictions or the measurements.}.

As a general trend we see in Fig. \ref{fig:Qm_signif_alpha} that all predictions differ more significantly from the measurements for smaller triangles.
This can be explained by the interplay of two effects. On one hand, terms in the perturbative expansion of $Q_m$ beyond leading order,
which are neglected in our $Q_m$ model, contribute stronger at smaller scales.
This explanation is consistent with the fact that the deviations are less significant at higher redshift and also when $Q_m^{PT}$ is computed
from the non-linear, instead of the linear power spectrum.  On the other hand, the signal-to-noise ratio is higher at small scales (see bottom panel of
Fig. \ref{fig:Qm_signifsimpt_r1r2r3}). Note that the latter is specific to the joint volume of our $49$ realisations of roughly $\sim100 \ (h^{-1} \text{Gpc})^{3}$.
For the smaller volumes of current and near future galaxy surveys we expect the model to deviate less significantly because of larger measurement errors.

Predictions from the de-wiggled power spectrum are very similar to those from the linear power spectrum for smaller triangles (e.g. $(r_1,r_2)=(64,32)$ \mpc{},
$\alpha \lesssim 120\degree$) and agree well with those from the non-linear power spectrum for large triangles (i.e. $(r_1,r_2)=(80,48)$ \mpc{}, $\alpha \gtrsim 90\degree$).
The latter finding indicates that for the tree-level calculation of the 3PCF in configurations space, implementing resummations over large-scale displacements by using the
de-wiggled power spectrum has almost the same effect as using the  non-linear spectrum from the simulation.
For both cases the predictions are in $1\sigma$ agreement with the measurements at the BAO scale, while using the linear spectrum leads to
$2-3\sigma$ deviations.
%
For the remainder of our analysis we will use predictions based on the non-linear power spectrum, as they show the best overall agreement with the
measurements in Fig. \ref{fig:Qm_signif_alpha}.

\begin{figure}
  \centering
   \includegraphics[width=70mm, angle = 270]{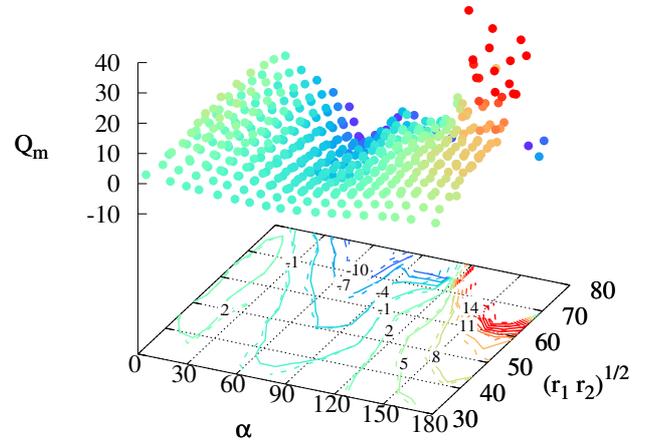}
   \caption{$Q_m$ measurements at $z=0.5$ versus the mean triangle opening angle per bin $\alpha$ and triangle scale $(r_1r_2)^{1/2}$.
   Dots and solid contour lines show mean results from $49$ simulations. Dashed contour lines show tree-level predictions
   based on the non-linear power spectrum. The colours indicate the amplitude of $Q_m$.}
\label{fig:Qm_alpha_r1r2}
\end{figure}

A convenient way to show results for all triangles in our analysis is to display them for a given opening angle $\alpha$
versus the triangle size, here defined as $\sqrt{r_1r_2}$. As an example we show the measurements of $Q_m$ in
Fig. \ref{fig:Qm_alpha_r1r2}. This figure demonstrates the strong increase of the u-shape of $Q_m(\alpha)$ with the triangle scale.
The minimum lies between $60$ and $90$ \degree. Measurements for $\alpha \gtrsim 120$ \degree and $\sqrt{r_1r_2} \gtrsim 50$
are dominated by noise.

\begin{figure}
   \includegraphics[width=120mm, angle = 270]{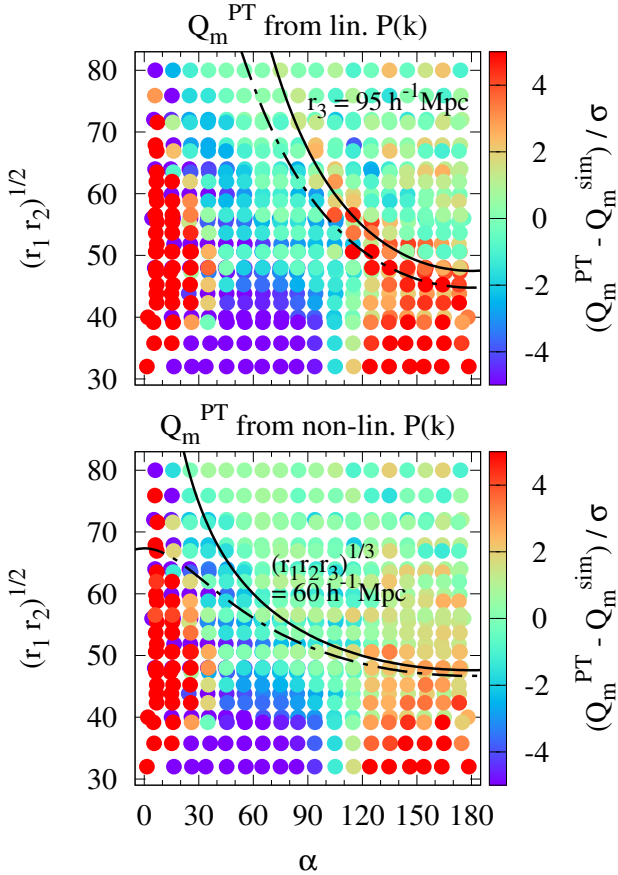}
	\caption{Significance of the deviations between $Q_m$ measurements and tree-level predictions
	versus the mean triangle opening angle per bin $\alpha$ and the triangle scale $(r_1r_2)^{1/2}$ in \mpc{} at redshift $z=0.5$.
	The predictions are derived from the linear and measured (non-linear) power spectrum (top and bottom panel
	respectively).
	Black lines in the top panel trace the BAO feature $(r_1r_2)^{1/2}(\alpha)$ for $r_3=95$\mpc{}.
	In the bottom panel black lines indicate the triangle scale at which the model fails at $2\sigma$,
	$(r_1r_2r_3)^{1/3} \simeq 60$\mpc{}. In both cases solid and dashaed-dotted lines correspond to
	triangle configurations with of $r_2 / r_1 = 1.0$ and $0.5$ respectively.}
	\label{fig:Qm_signifsimpt_alpha_r1r2_proj}
\end{figure}

The significance of the deviations between $Q_m$ model predictions and measurements
are shown for redshift $z=0.5$ versus $\alpha$ and $\sqrt{r_1r_2}$ in Fig. \ref{fig:Qm_signifsimpt_alpha_r1r2_proj}.
We find that $Q_m^{PT}$ is below the measurements for opening angles between roughly $30-90$ \degree for triangles
with $30 \lesssim \sqrt{r_1r_2} \lesssim 50$ \mpc{}. For smaller and larger opening angles the predictions tend to lie above the measurements.
Similar results based on simulations with different cosmologies have been reported in the literature
(see for instance \citet{bargaz} or \citet{HBG15a}, who use the same algorithms for the $Q_m$
predictions and measurements as employed in this study). \citet{Scoccimarro98} showed that such deviations
can be explained by higher order contributions as they reduce when the predictions are developed to next to
leading order, including 1-loop terms \citep[see also][]{Sefusatti10}.
As in Fig. \ref{fig:Qm_signif_alpha} one can see in the top panel of Fig. \ref{fig:Qm_signifsimpt_alpha_r1r2_proj} that using
the linear power spectrum leads to strong deviations between predictions and measurements, in particular around the
BAO peak, which are apparent as a red banana-shaped feature.
This BAO feature follows roughly triangles with $r_3 \sim 95$ \mpc{}, which are marked in the top panel as black lines.
The deviations strongly reduce when the predictions are computed from the non-linear power spectrum for triangles scales
$\sqrt{r_1r_2} \gtrsim 50$ \mpc{}  and $\alpha \gtrsim 30$ to roughly $1\sigma$.

Defining the overall triangle size as $(r_1r_2r_3)^{1/3}$, we find that the deviations converge
to $2 \sigma$ at $(r_1r_2r_3)^{1/3} \gtrsim 60$ ($80$) \mpc{}, when using the non-linear (linear) power
spectrum (Fig. \ref{fig:Qm_signifsimpt_r1r2r3}). Triangles with $(r_1r_2r_3)^{1/3}=60$ \mpc{}
are therefore marked by black lines in the bottom panel of Fig. \ref{fig:Qm_signifsimpt_alpha_r1r2_proj}).
%
\begin{figure}
  \centering
   \includegraphics[width=75mm, angle = 270]{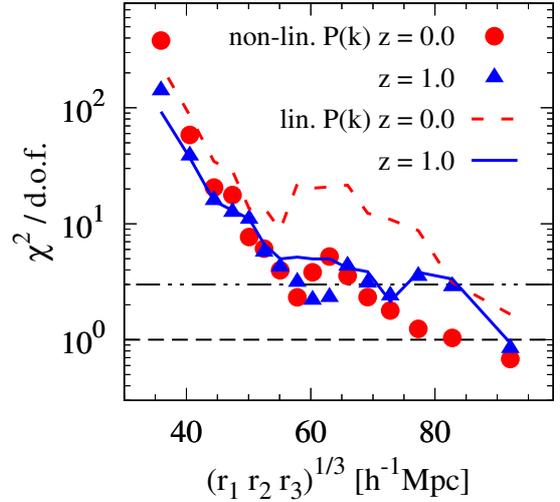}
   \caption{$\chi^2 / d.o.f.$, quantifying the significance of the deviation between mean $Q_m$, measured in the simulations and
   tree-level predictions versus the mean triangle size $(r_1r_2r_3)^{1/3}$ per bin. Lines and symbols show results
   for $Q_m$ predictions derived from the linear and non-linear power spectrum respectively at the redshifts $z=0.0$ and $z=0.5$.}
\label{fig:Qm_chisq}
\end{figure}
The normalised $Q_m$ covariance matrix, shown in Fig. \ref{fig:Q_cov}, reveals that the $Q_m$
measurements for different triangles are correlated with each other. Hence, we compute an SVD estimate of the $\chi^2$
in bins of $(r_1r_2r_3)^{1/3}$ to quantify the deviation between measurements and predictions, taking the covariance into account,
as described in Section \ref{subsec:3pc_errorrs}. Each \rtr{} bin includes measurements from $30$ triangles, while
we tested that our results change only weakly, when using $20$ and $40$ triangles per bin and do not affect our conclusions.

In Fig. \ref{fig:Qm_chisq},
we find $\chi^2/d.o.f.$ values between $10$ and $100$ for $(r_1r_2r_3)^{1/3} \lesssim 50$ \mpc{}
at $z=1.0$, where the degree of freedom ($d.o.f.$) is the number of singular values used for the $\chi^2$ estimation.
At $z=0.0$ the $\chi^2/d.o.f.$
values are higher at small scales, indicating that $Q_m$ predictions agree better with measurements at higher redshifts.
At $(r_1r_2r_3)^{1/3}  \gtrsim 60$ \mpc{} the $\chi^2/d.o.f.$ values are roughly constant, taking values between $0.6-4$.
An exception are the high values for the $Q_m$ model from the linear power spectrum at $z=0.0$, whereas using the non-linear
and linear power spectra lead to similar results at $z=1.0$. These results indicated that non-linear contributions have
a significant effect in $Q_m$ at small scales and low redshift and can partly be taken into account in the $Q_m$ predictions
by using the non-linear power spectrum.

Note that the differences between results at different redshifts do not only result from different model performance,
but also from differences in the covariances and modes selected for the $\chi^2$ computation.
Since these quantities are sensitive to noise we will not enter a detailed discussion.

\subsection{$\Delta Q$}\label{subsec:accuracy_dQ}

In this subsection we test how well the higher-order contributions to the halo 3PCF are described by the
quadratic $c_2+g_2Q_{nloc}$ term, which appears in equation (\ref{eq:Qh_model}) and (\ref{eq:Qx_model}). Following \citet{BHG15},
we obtain these higher-order contributions from the measurements by subtracting the halo-matter cross-correlation from the halo
auto-correlation,
\begin{equation}
	\Delta Q \equiv Q_h - Q_\times.
	\label{eq:dQ_def}
\end{equation}
This subtraction leads to a cancellation of the linear $Q_m/b_1$ term in $Q_h$ and $Q_\times$ and hence isolates the higher-order terms.
The aforementioned quadratic term correspond to the leading order perturbative approximation of $\Delta Q$,
which follows from inserting the corresponding leading order approximations for $Q_h$ and $Q_\times$ from
equation (\ref{eq:Qh_model}) and (\ref{eq:Qx_model}) into equation (\ref{eq:dQ_def}), i.e.
\begin{equation}
	\Delta Q \simeq \frac{2}{3 b_1}(c_2 + g_2 Q_{nloc}).
	\label{eq:dQ_model}
\end{equation}
The relation above allows us to test on one hand the accuracy of the quadratic model for the higher order terms in $Q_h$ and $Q_\times$,
independently of inaccuracies in the $Q_m$ modeling, which we studied previously in Section \ref{subsec:accuracy_Qm}.
On the other hand, we test simultaneously if the bias parameters, which we adopt from the Fourier space measurements
of \css{}, also describe the clustering statistics in configurations space. Regarding the latter case we employ three sets
of bias parameters to which we refer to as local, non-local and \bdsqfix{} bias model, as described in Table \ref{table:bias_models},
Section \ref{sec:bias_models}.

\begin{figure}
  \centering
   \includegraphics[width=120mm, angle = 270]{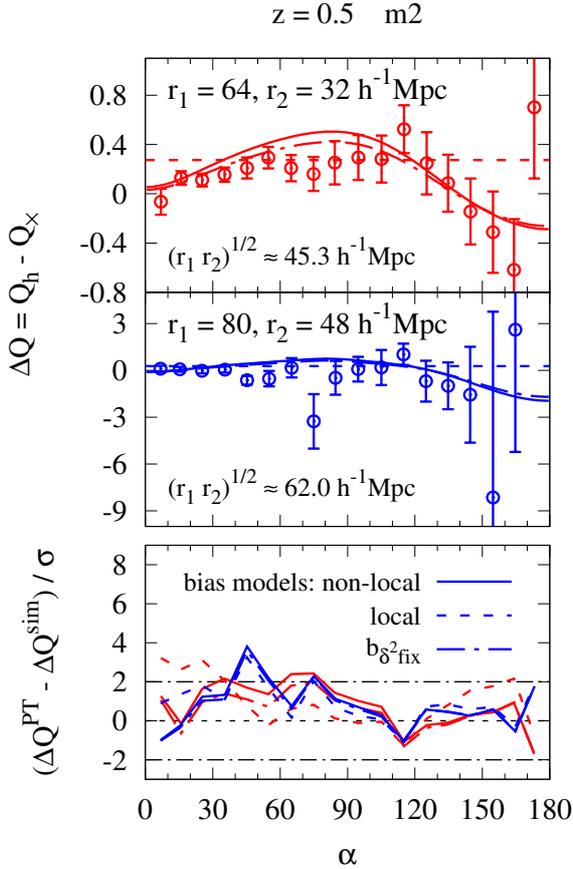}
   \caption{Top panels: $\Delta Q$ versus the mean triangle opening angle per bin $\alpha \equiv \arccos( \hat{\bf r}_1 \cdot \hat{\bf r}_2)$,
   measured at $z=0.5$ for the mass sample m2.
   Dashed and solid lines show tree-level predictions from the local and non-local bias model respectively, using the non-linear power spectrum and bias
   parameters measured in Fourier space by \css{} in equation (\ref{eq:dQ_model}). Bottom panel: the significance of the deviation between
   model predictions and measurements.}
   \label{fig:dQ_alpha}
\end{figure}

The corresponding model predictions for $\Delta Q$ are compared to the measurements at different triangle
opening angles in Fig. \ref{fig:dQ_alpha}. For this comparison we use the halo sample m2 at redshift
$z=0.5$ (defined in Table \ref{table:nyu_halo_masses}) and the same triangle configurations as for the
$Q_m$ in Fig. \ref{fig:Qm_alpha}.
The $\Delta Q$ measurements in Fig. \ref{fig:dQ_alpha} show a clear dependence on the triangle
opening angle $\alpha$ for the small $(r_1,r_2) = (64,32)$ \mpc{} triangle configuration. This finding
contrasts the local bias model prediction of a constant $\Delta Q= 2c_2 / 3 b_1$. However, at intermediate angles 
($60\degree \lesssim \alpha \lesssim 120 \degree$) the local model predictions are in better agreement
with the measurements than predictions from the non-local model. This result indicates that neglected higher order
terms might compensate the quadratic non-local contribution.

\begin{figure*}
  \centering
   \includegraphics[width=70mm, angle = 270]{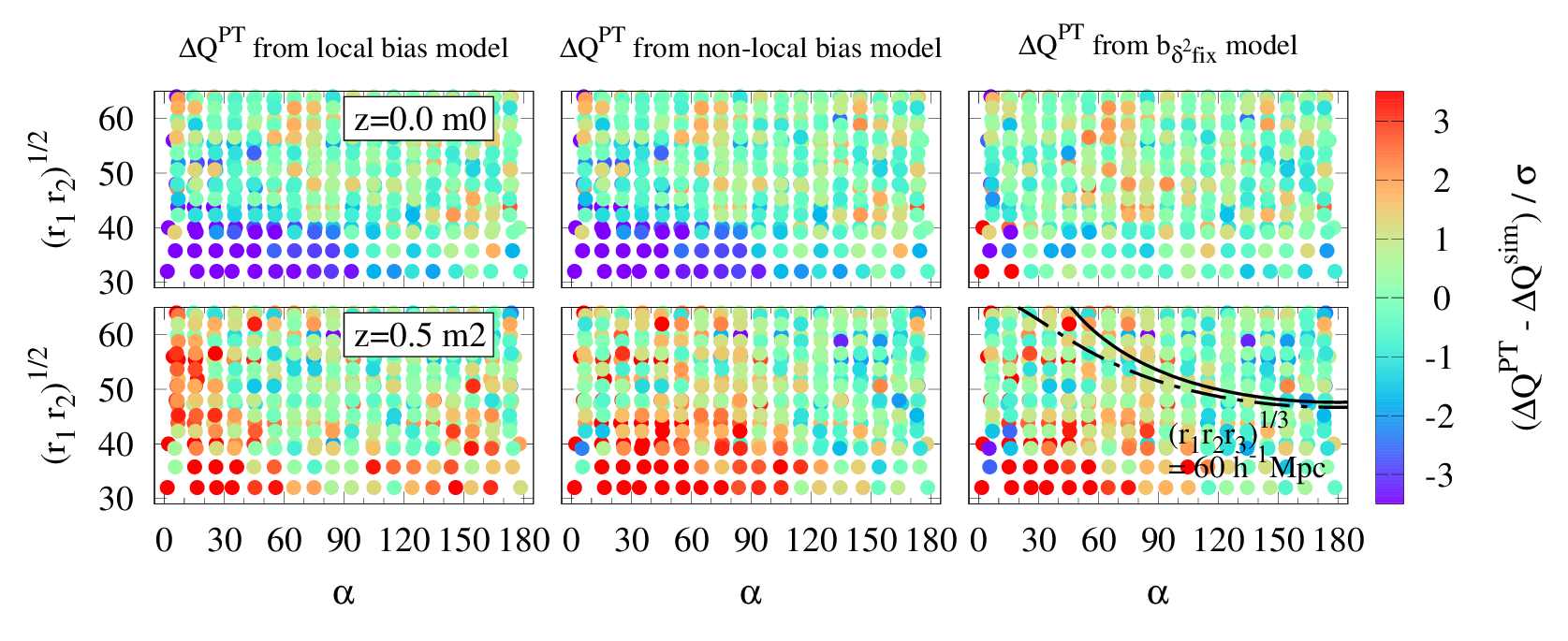}
   \caption{Significance of the deviations between predictions for $\Delta Q$
	and measurements for the mass samples m0 at $z=0.0$ and m2 at $z=0.5$ (top and bottom panels respectively).
	The results are shown versus the mean triangle opening angles per bin $\alpha$ and triangle scales $(r_1r_2)^{1/2}$ in \mpc{}. 
	The predictions are derived from the local, the non-local and the \bdsqfix bias model (see Table \ref{table:bias_models}),
	with bias parameters measured in Fourier space by \css.
   Solid and dashed black lines in the bottom right panel show $(r_1r_2)^{1/2}(\alpha)$ for $(r_1r_2r_3)^{1/3}=60$\mpc{} and triangle
   configurations of $r_1/r_2 = 0.5$ and $1.0$ respectively.}
 \label{fig:dQ_signifsimpt_alpha_r1r2_proj}
\end{figure*}

Similar trends are apparent for the larger ($80,48$) \mpc{} triangle configuration, while here the large measurement errors 
lead to a similar significance of the different model deviations (see bottom panel of Fig. \ref{fig:dQ_alpha}).

Note that for the presented results we computed $Q_{nloc}$ in equation (\ref{eq:dQ_model}) from the non-linear
power spectrum, which was measured in the simulation. This is motivated by the fact that the $Q_m$ model
performs better in that case (see Section \ref{subsec:accuracy_Qm}).  However, using $Q_{nloc}$ predictions
from the linear power spectrum delivers very similar result and does not affect the conclusions drawn above.

Extending the comparison between models and measurements to all triangles in our analysis, we show
in Fig. \ref{fig:dQ_signifsimpt_alpha_r1r2_proj} the significance of the deviations  between $\Delta Q$ measurements and model predictions
versus the triangle opening angle and scale $\sqrt{r_1r_2}$ (analogously to Fig. \ref{fig:Qm_signifsimpt_alpha_r1r2_proj}).
We use again the mass sample m2 at $z=0.5$ (with $b_1=3.29$) and show in addition also results for the sample m0 at $z=0.0$ (with $b_1=1.43$)
to explore how differences in the bias effect the model performance.
For the highly biased sample m2 at $z=0.5$ (bottom panel of Fig. \ref{fig:dQ_signifsimpt_alpha_r1r2_proj}) the results line up with those
for the two single triangle configurations, shown Fig. \ref{fig:dQ_alpha}.
For small triangles with $\sqrt{r_1r_2} \lesssim 40$\mpc{} and triangle opening angles in the range of $60\degree-120\degree$ the
local bias model is in better agreement with the measurements than the non-local model. Overall both models tend to overpredict the
measurements at small triangle scales. The results for the \bdsqfix{} model are very similar to those from the non-local model.
This is also the case when the latter is based on the $b_2(b_1)$ relation from \citet{Lazeyras15}.

These findings differ from those of the low biased sample m0 at $z=0.0$ (shown in top panel of Fig. \ref{fig:dQ_signifsimpt_alpha_r1r2_proj})
in three aspects. The first aspect is that the local and non-local model tend to underpredict the measurements for $\sqrt{r_1r_2} \lesssim 40$ \ \mpc{}.
The second aspect is that for the low biased sample the local and non-local model perform equally well. This can be expected,
since the non-local bias, measured by \css{} is close to zero in that case. The third aspect is that the \bdsqfix{} model differs from
non-local model. In fact, it agrees better with the measurements than the other models. One interpretation of this result could be
that the $c_2(b_1)$ and $g_2(b_1)$ relation is more accurate than the Fourier space measurements of the bias parameters from \css{}.
Alternatively one might conclude that inaccuracies of the \bdsqfix{} model compensate the neglected higher-order terms in the $\Delta Q$
model in equation (\ref{eq:dQ_model}), leading to a good agreement with the measurements by accident. To clarify this point one could
repeat the exercise, using a model for $\Delta Q$ which is developed beyond second order. For a possible application of the $c_2(b_1)$
and $g_2(b_1)$ relations of the \bdsqfix{} model in observations it would be interesting to test the dependence of our results
on the cosmological parameters used.
For bias measurements in observations it is also interesting to note that deviations between measurements in our
$\sim100 \ (h^{-1} \text{Gpc})^{3}$ volume and model predictions become insignificant for $\sqrt{r_1r_2} \gtrsim 40$\mpc{} as the
measurement errors increase with scale.

As for $Q_m$ we find an overall convergence of the deviation between measurements and
predictions for triangles with \rtr{} $\gtrsim 60$ \mpc{} in Fig. \ref{fig:dQ_signifsimpt_r1r2r3},
which are marked in Fig. \ref{fig:dQ_signifsimpt_alpha_r1r2_proj} with black lines.
%
\begin{figure}
  \centering
   \includegraphics[width=80mm, angle = 270]{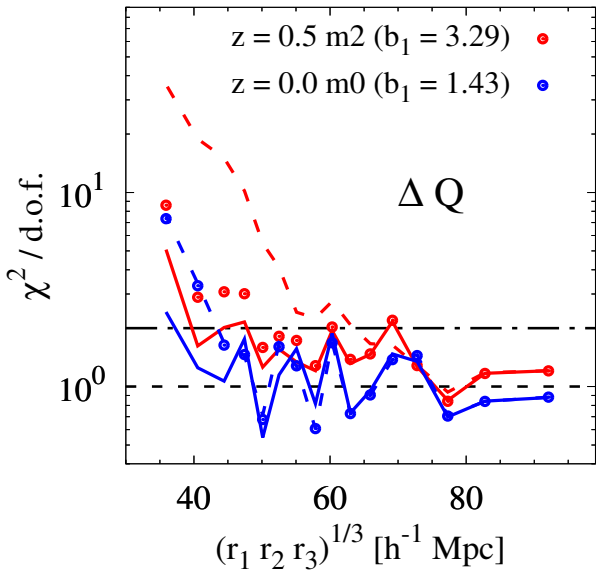}
   \caption{$\chi^2/d.o.f.$, quantifying the significance of the deviation between the mean 
   $\Delta Q$, measured in the simulations and predictions based on Fourier space 
   bias parameters. Results are shown versus the mean triangle size \rtr{} per bin.
   Dots are results based on the predictions from the non-local bias model.
   Results from the local and the $b_{\delta^2fix}$ bias model (see Table \ref{table:bias_models})
   are shown as dashed and solid lines respectively.
   Note that for the low biased mass sample m0 at redshift $z=0.0$
   the results for the local and non-local bias are very similar
   since the non-local bias contribution is very small.}
\label{fig:dQ_chisq}
\end{figure}
We quantify these deviations again by computing the $\chi^2$ via SVD, taking into account the
covariance between measurements at different scales in \rtr{} bins with $30$ triangles.
Note that the $\Delta Q$ covariance is typically dominated by shot noise, coming from the $Q_h$
contribution, which can be seen in Fig. \ref{fig:Q_cov}.
The results, shown in Fig. \ref{fig:dQ_chisq} are in line with our finding from Fig. {\ref{fig:dQ_signifsimpt_r1r2r3} as results
converge to $\chi^2/d.o.f$ values around unity. The highly biased sample shows
larger overall deviations between measurements and predictions, in particular for the non-local model at
\rtr{} $\lesssim 60$\mpc{}. Results for the \bdsqfix{} model are similar to those from the non-local model
at large scales, while at small scales the former performs better as its $\chi^2/d.o.f.$ values are lower.

\begin{figure*}
  \centering
   \includegraphics[width=120mm, angle = 270]{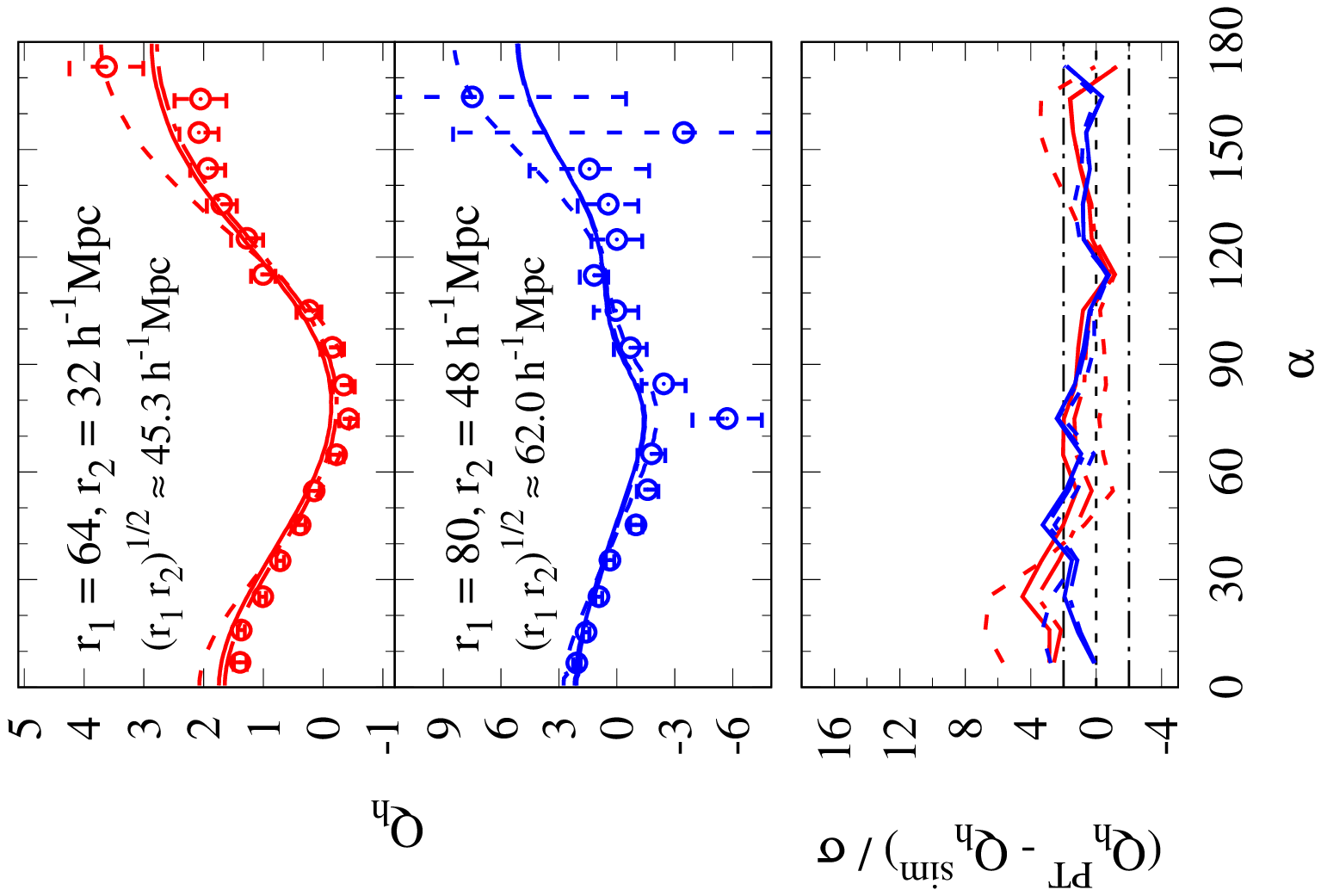}
   \includegraphics[width=120mm, angle = 270]{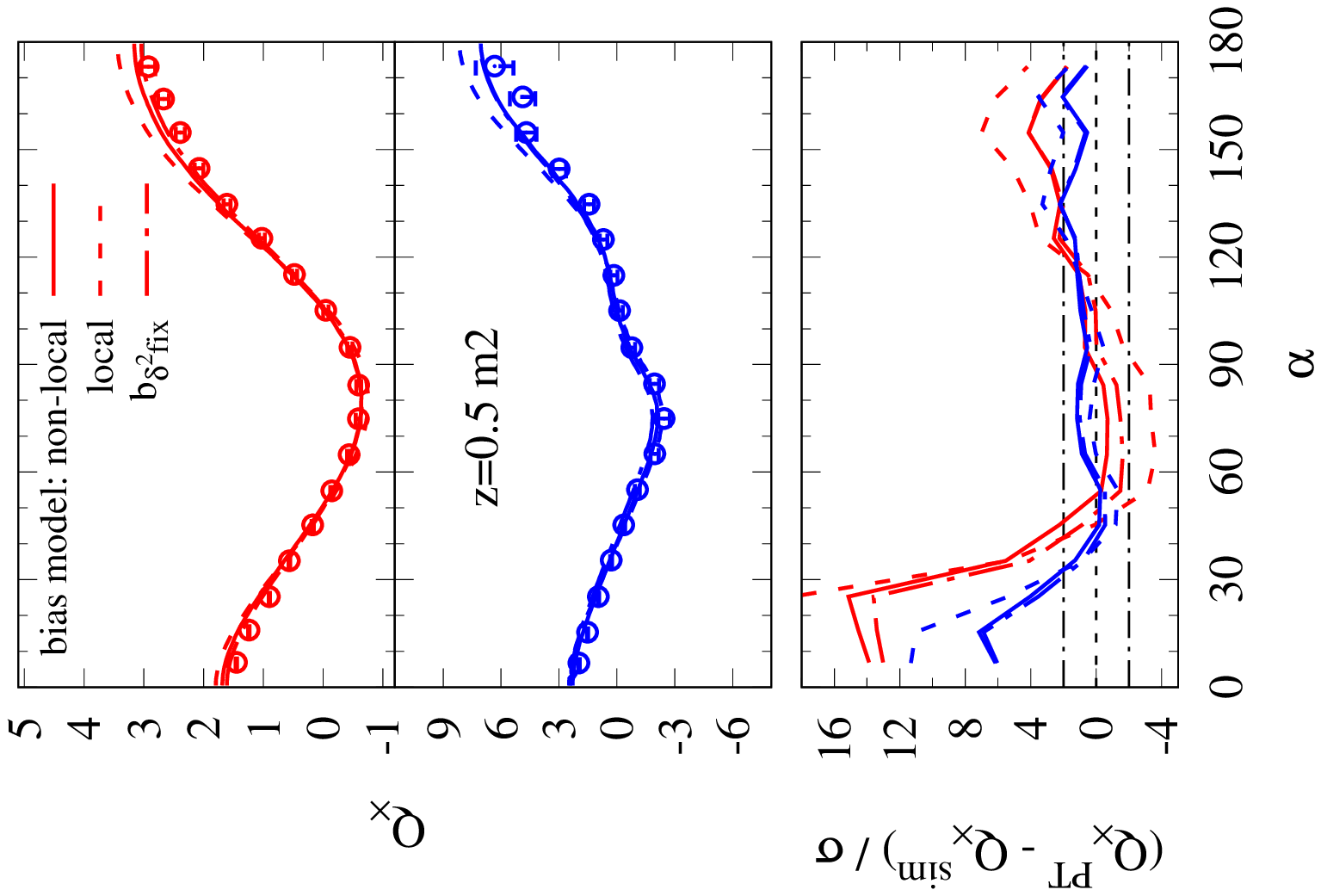}
    \caption{Left: top panels show the reduced halo 3PCF, $Q_h$, for triangles with fixed legs $r_1$ and $r_2$
   (size is indicated in \mpc{}) versus the mean triangle opening angle per bin, $\alpha \equiv \arccos( \hat{\bf r}_1 \cdot \hat{\bf r}_2)$.
   Symbols show mean measurements from $49$ simulations with $1\sigma$ errors for the mass sample m2 at redshift $z=0.5$. Lines show predictions
   from equation (\ref{eq:Qh_model}), using the non-linear power spectrum and the bias models from Table \ref{table:bias_models} with bias
   parameters measured by \css{} in Fourier space. The bottom panel shows the significance of the deviations between model predictions and measurements.
   Right: analogous results for the reduced three-point halo-matter cross-correlations, while predictions are derived from equation (\ref{eq:Qx_model}).
 }
\label{fig:Qhx_alpha}
\end{figure*}

\subsection{$Q_h$ and $Q_{\times}$}\label{subsec:accuracy_QhQx}

After validating the linear and quadratic components for the $Q_h$ and $Q_\times$ models separately
in Section \ref{subsec:accuracy_Qm} and \ref{subsec:accuracy_dQ} we now compare the full models,
given by equation (\ref{eq:Qh_model}) and (\ref{eq:Qx_model}) with the measurements in our simulations.
As for $\Delta Q$ we focus on model predictions, which are based on the non-linear power spectrum
and start the analysis by 
showing $Q_h$ and $Q_\times$, measured in the halo sample m2 at $z=0.5$, for triangles with
fixed legs of $(r_1,r_2)=(64,32)$ and $(80,48)$\mpc{} versus the triangle opening angle $\alpha$
in Fig. \ref{fig:Qhx_alpha}.
We find that the models for both, $Q_h$ and $Q_\times$ tend to overpredict the measurements,
which lines up with our corresponding results for $\Delta Q$ in Fig. \ref{fig:dQ_alpha}.
An exception of this trend are $Q_h$ results from the small triangle configuration
with $60 \lesssim \alpha \lesssim 90$. This indicates that the neglected terms in the perturbative model beyond
leading order affect $Q_h$ and $Q_\times$ differently. Again, the model predictions based on the local bias model
show the strongest deviations from the measurements, in particular for collapsed and relaxed triangles.
This explains why neglecting the non-local term leads to an overestimation of the bias, when fitting
$Q_h$ or $Q_\times$ model predictions to measurements (see \css{}). For such a fit one would choose a
higher $b_1$, since this would flatten the curve and deliver the measured shape. The overall amplitude
can then be adjusted by varying $c_2$ (see equation (\ref{eq:Qh_model})). Such fits of the local model
are in fact in very good agreement with the measurements. However, the linear bias is too high
\citep[e.g.][]{M&G11, BHG15}. Note that the linear bias measurements based on the local bias model would
be too low instead of too high when using the 3PCF or the bispectrum, as explained by \css{}.

The best agreement between the $Q_h$ and $Q_\times$ measurements and the corresponding models
occurs at large opening angles (hence large triangles) when using the non-local bias model ($1-2\sigma$).
This scale dependence can be expected since errors increase and higher order contributions decrease with the scale.
Results based on the \bdsqfix{} model are again very similar to those from the non-local model.

Interestingly the deviations between the model predictions and measurements are less significant for $Q_h$ than for $Q_\times$,
despite the fact that the neglected terms beyond leading order should have a higher contribution to $Q_h$ and
therefore lead to stronger deviations from the model. However, the errors on $Q_h$ are more strongly affected by
shot-noise than those for $Q_\times$ ($\sigma^2_{Q_h} \sim n_h^3$, $\sigma^2_{Q_\times} \sim n_h$, where $n_h$
is the halo number density). This means that for observations with similar or larger errors than our measurements,
a development of the $Q_h$ model beyond leading order might only lead to a marginal improvement of the model performance.

\begin{figure*}
  \centering
   \includegraphics[width=70mm, angle = 270]{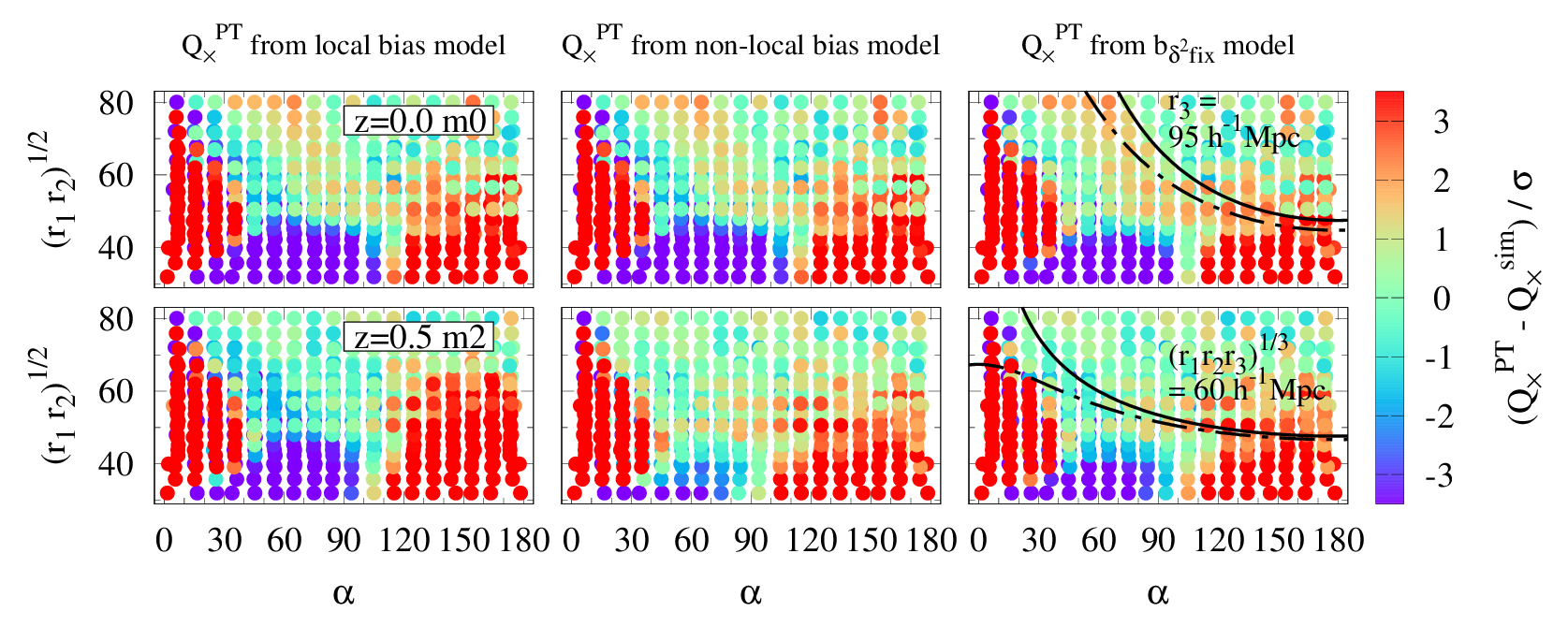}
   \includegraphics[width=70mm, angle = 270]{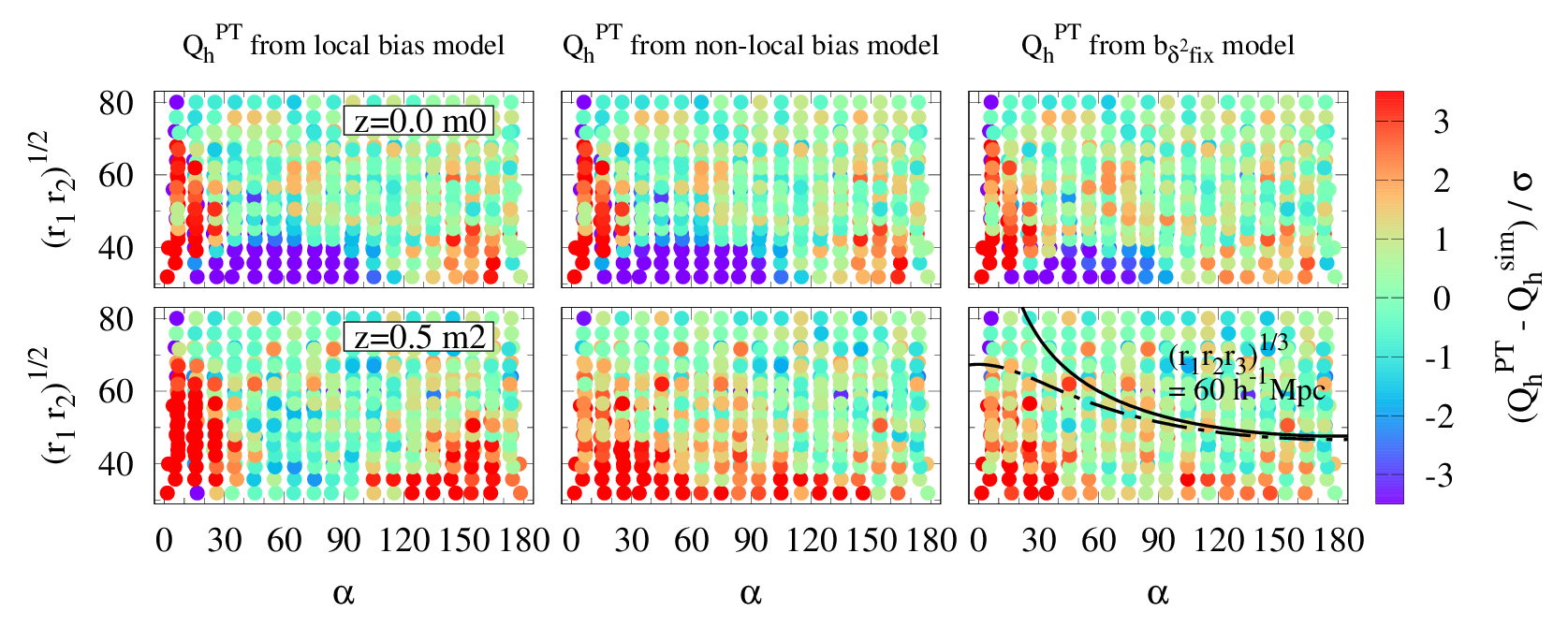}
   \caption{Significance of the deviations between predictions for $Q_\times$ and $Q_h$
   	and measurements for the mass samples m0 at $z=0.0$ and m2 at $z=0.5$ (top, bottom panels respectively).
	The results are shown versus the mean triangle opening angles per bin, $\alpha$, and triangle scales $(r_1r_2)^{1/2}$ in \mpc{}. 
	The predictions are derived from the bias models described in Table \ref{table:bias_models}, with bias parameters
	measured in Fourier space by \css. Solid and dashed black lines in the bottom right panel show
	$(r_1r_2)^{1/2}(\alpha)$ for $(r_1r_2r_3)^{1/3}=60$\mpc{} and triangle configurations of $r_1/r_2 = 0.5$ and $1.0$ respectively.}
\label{fig:Qhx_signifsimpt_alpha_r1r2_proj}
\end{figure*}

In Fig. \ref{fig:Qhx_signifsimpt_alpha_r1r2_proj} we show the comparison between $Q_\times$ and $Q_h$ models
and measurements for all triangles, displaying them for different scales $\sqrt{r_1r_2}$ versus the triangle opening angle
(as in Fig. \ref{fig:dQ_signifsimpt_alpha_r1r2_proj}). Results are shown for the low biased sample m0 at $z=0.0$
and the highly biased sample m2 at $z=0.5$. The latter confirm the trends from from Fig. \ref{fig:Qhx_alpha}.
In particular for small triangles  ($\sqrt{r_1r_2} \lesssim 40$\mpc{}) the $Q_\times$ and $Q_h$ models overpredict the
measurements for collapsed and relaxed triangles and underpredict them for triangles with $60 \lesssim \alpha \lesssim 90$.
The $Q_h$ results for m2 at $z=0.5$ are again an exception. In that case the local model is in better agreement
with the measurements than the non-local model, which is consistent with the $\Delta Q$ results for this sample
and might be attributed to a compensation of quadratic non-local and neglected higher-order terms, as mentioned
in the discussion of Fig. \ref{fig:dQ_alpha} in Section \ref{subsec:accuracy_dQ}. Note that this compensation
is shown here to occur for one particular halo sample, while this is not the case for other samples (not shown here).

Overall the results from the non-local bias model are in better agreement with the measurements for the sample
m2 at $z=0.5$ than the local model at large triangles scales ($\sqrt{r_1r_2} \gtrsim 40$ \mpc{} or $\alpha \gtrsim 90$)
and are consistent with those from the \bdsqfix{} model.
For the low biased sample m0 at $z=0.0$ all bias models deliver similar results, since the non-local bias contribution is very weak.
For some triangles we find an increased significance of the deviations for that sample, compared to the m2 sample at $z=0.5$,
presumably because the shot noise error contribution is decreased due to the higher halo density. In the case of $Q_\times$, where
the shot noise errors are the lowest, the deviations follow the BAO feature, which we saw already in the $Q_m$ model validation
(Fig. \ref{fig:dQ_signifsimpt_alpha_r1r2_proj}).
Using model predictions based on the linear power spectrum, we find a significant increase of the deviations
in the case of $Q_\times$ for both samples (not shown here).
This indicates that neglected terms in the $Q_m$ model, in the bias model or both affect the halo 3PCF,
even at very large triangle scales. However, for the auto-correlation $Q_h$ at $\sqrt{r_1r_2} \gtrsim 40$ \mpc{}
their contribution seems to be small compared to the measurement errors, as we find a similarly significant
deviations for different power spectra and halo samples. This will in particular be also the case for
the smaller volumes, covered by galaxy surveys, for which the measurement errors can be expected to be larger.

The deviations between non-local bias model predictions
and measurements converge to values of $\lesssim 2 \sigma$  \rtr{}$\gtrsim 60$\mpc,
(marked in Fig. \ref{fig:Qhx_signifsimpt_alpha_r1r2_proj} as black lines)
for both, $Q_\times$ and $Q_h$, as shown in Fig. \ref{fig:Qx_signifsimpt_r1r2r3} and \ref{fig:Qh_signifsimpt_r1r2r3}.
This is consistent with our corresponding results for $Q_m$ and $\Delta Q$.

\begin{figure*}
  \centering
   \includegraphics[width=120mm, angle = 270]{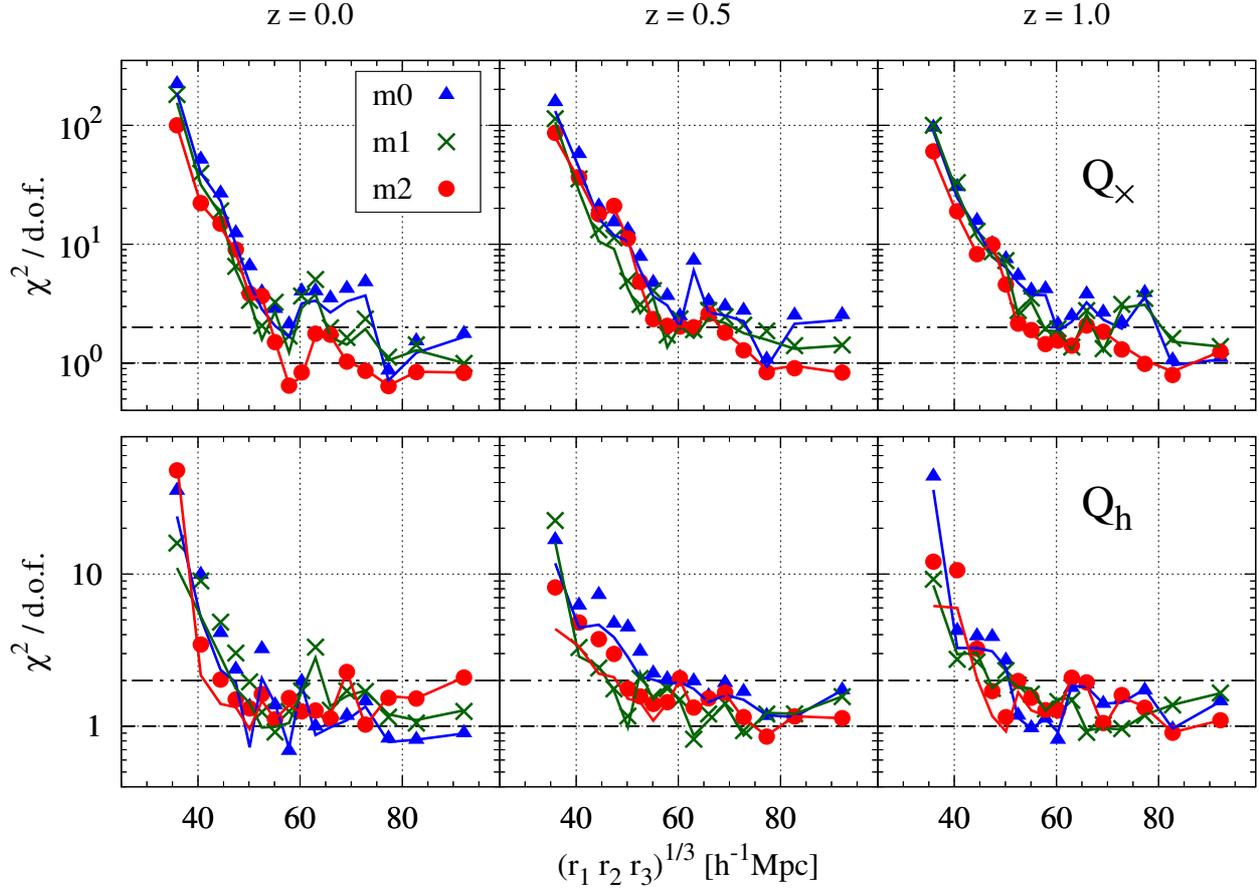}
   \caption{Top: $\chi^2$ per degree of freedom (d.o.f.), quantifying the difference between
   the $Q_\times$ measurements for the mass samples m1, m2, m3 (defined in Table \ref{table:nyu_halo_masses})
   and the corresponding predictions, based on the non-linear power spectrum at different redshifts $z$.
   Symbols show results for the non-local bias model, while lines show results using analytical relations between the
   linear and non-linear bias parameters (\bdsqfix{} model, see Table \ref{table:bias_models}).
   The bias parameters were measured by \css{} in Fourier space. Bottom: same as top panel, but for $Q_h$.}
\label{fig:QhQx_chisq}
\end{figure*}

As in the case of $Q_m$ and $\Delta Q$, measurements of $Q_h$ and $Q_\times$ from different triangles
are covariant (see Fig. \ref{fig:Q_cov}). Quantifying the significance of deviations between model predictions
and measurements, we show in Fig. \ref{fig:QhQx_chisq} the $\chi^2/d.o.f.$ in bins of \rtr{} for all three mass samples and redshifts.
Each bin contains measurements from $30$ triangles and the $\chi^2$ values have been computed via SVD (see Section \ref{subsec:3pc_errorrs})
using only the dominant modes, as for the $Q_m$ and $\Delta Q$ analyses from Fig. \ref{fig:Qm_chisq} and \ref{fig:dQ_chisq}.
We also tested that our results are not affected by the chosen number of triangles per bin.
The results confirm the convergence of the deviations to $1-2 \sigma$ for \rtr{} $\gtrsim 60$ \mpc{}.
However, they show strong variations for different scales, which might result from noise in our covariance estimation
from only $49$ realisations.
Overall, the $\chi^2/d.o.f.$ values for $Q_\times$ are higher than those for $Q_h$, presumably because of the higher
signal-to-noise ratio of the measurements. Even at large scales above $60$ \mpc{} we find $\chi^2/d.o.f.\simeq 5$ values.
They might be explained by non-linearities around the BAO feature, which are not fully captured in our leading order
perturbative model (see Fig. \ref{fig:Qhx_signifsimpt_alpha_r1r2_proj}).
For $Q_\times$ the $\chi^2/d.o.f.$ values are lower at high redshift and higher mass samples. The latter result might be explained by
larger shot-noise errors on the high mass samples and agrees with the results from Fig. \ref{fig:Qx_signifsimpt_r1r2r3}
and \ref{fig:Qh_signifsimpt_r1r2r3}.
Smaller deviations at high redshifts might result from a smaller impact of next to leading order terms in the $Q_\times$ model,
which we neglect in our analysis. We do not see a clear dependence of the results on mass and redshift for $Q_h$,
possibly because  of the low signal-to noise ratio. It is interesting to note that the $\chi^2/d.o.f.$ values for the \bdsqfix{}
model are in very good agreement with those from the non-local model for highly biased sample (high halo mass and redshift).
For samples with low bias (low mass, low redshift) the $\chi^2/d.o.f.$ values for the \bdsqfix{} are even smaller than
those for the non-local model. The latter finding is consistent with our model comparison for $\Delta Q$.

Our comparison between $\chi^2/d.o.f.$ values for local and non-local model predictions in Fig. \ref{fig:QhQx_chisq_locmod}
demonstrates that setting the non-local term in the prediction to zero leads to higher deviations from $Q_\times$ measurements
for highly biased samples. The effect is also apparent for $Q_h$, even for \rtr{} $>60$ \mpc{}, while in that case the $\chi^2/d.o.f.$ values
are lower, presumably due to larger errors on the measurements. Again these results confirm those for $\Delta Q$, shown in Fig. \ref{fig:dQ_chisq}. 

\begin{figure}
  \centering
   \includegraphics[width=120mm, angle = 270]{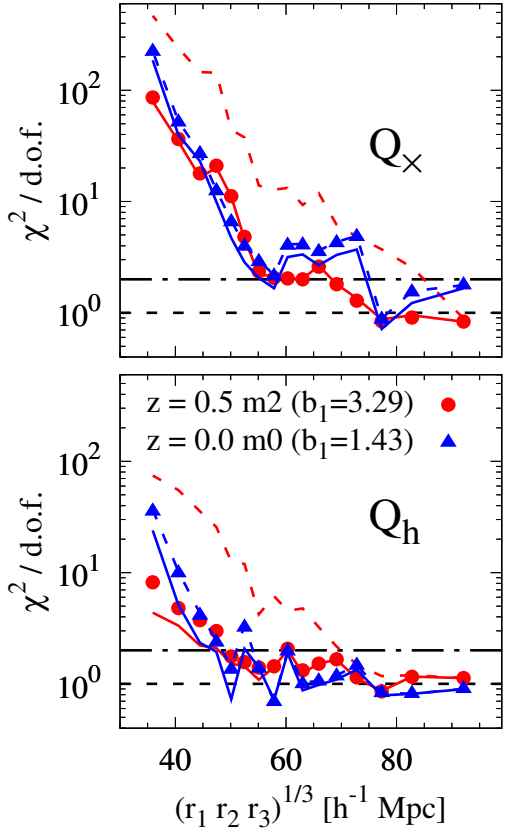}
      \caption{The figure shows the same results for the non-local and \bdsqfix{} model
      as shown in Fig. \ref{fig:QhQx_chisq} for the mass sample m0 at $z=0.0$ m2 at $z=0.5$.
      In addition we show here results for the local model as dashed lines.}
\label{fig:QhQx_chisq_locmod}
\end{figure}

%% file: sections/conclusions.tex
\section{Summary and Conclusions}

The main result of this paper (summarized in Fig. \ref{fig:QhQx_chisq}) is an empirical
determination of the scales at which three-point halo correlations in configuration space
are consistent with the corresponding statistics in Fourier space, i.e. the bispectrum.
To this end, we measured the reduced three-point auto-correlation function of matter and halos,
as well as the reduced halo-matter three-point cross-correlation (which are referred to as $Q_m$, $Q_h$ and $Q_\times$ respectively)
in a set of $49$ cosmological simulations with a total volume of $\sim100 \ (h^{-1} \text{Gpc})^{3}$. The large volume provides small errors on the
measurements. At the same time we obtain rough estimates of the error covariances, which we analysed using singular value decomposition.
The $Q_h$ and $Q_\times$ measurements were compared to leading order perturbative models
(equation (\ref{eq:Qh_model}) and (\ref{eq:Qx_model})), which relate these statistics to  $Q_m$ via the linear,
quadratic and non-local bias parameters (referred to as $b_1$, $c_2$ and $g_2$ respectively).
For testing the consistency with results from Fourier space, we adopted bias parameters,
which were measured in the same set of simulations by \citet[][referred to as \css{}]{chan12} using the
same perturbative model of the halo-matter cross-bispectrum.

We adopted the bias parameters in three different ways. The first way is to simply employ the set of Fourier space parameter from \css.
The second set of parameters are identical to the first set, except for the non-local bias parameter $g_2$, which is set to zero in order to study
the contribution of the non-local terms to the $Q_h$ and $Q_\times$ predictions. For the third set we used the linear bias, measured by \css{}
from the halo-matter cross-power spectrum, while the quadratic bias is set by the (approximately) universal $c_2(b_1)$ relation from \citet[][]{HBG17}
and the non-local bias is predicted using the $g_2(b_1)$ relation from the local Lagrangian model, reducing the degrees of freedom in the bias model.
These three sets of bias parameters are referred to as non-local, local and \bdsqfix{} model respectively and are summarized in Table \ref{table:bias_models}.

Before predicting $Q_h$ and $Q_\times$ using the bias parameters, we first had to obtain the matter contribution $Q_m$ and the non-local contribution $Q_{nloc}$.
To remain closer to an analysis of observational data, where these quantities cannot be directly measured,
we modeled them from the linear, the linear de-wiggled and the non-linear power spectrum.
By comparing the $Q_h$ and $Q_\times$ predictions to measurements, we therefore did not only test if the bias parameters in Fourier space
describe the clustering in configurations space, but also simultaneously at which scales the perturbative model of the three-point correlation breaks down. 
We conducted this comparison in three steps. We first studied in Section \ref{subsec:accuracy_Qm} how well $Q_m$ measurements are described
by the leading order perturbative predictions from the different power spectra. Secondly, we investigate how well the higher-order contributions to $Q_h$
are described by the leading  (quadratic) order perturbative models, based on the Fourier space bias parameters. These contributions are
obtained from the measurements by the subtraction $\Delta Q \equiv Q_h - Q_\times$, as described in Section \ref{subsec:accuracy_dQ}.
Finally, we compare in Section \ref{subsec:accuracy_QhQx} the full predictions for for $Q_h$ and $Q_\times$ with the corresponding measurements.

Overall our results show that the deviations between the model predictions for $Q_m$, $Q_h$, $Q_\times$ and $\Delta Q$ and the corresponding
measurements depends on the triangle scale as well as on the triangle shape (characterised  by the triangle opening angle)
for which these statistics are studied.
The quantity $(r_1r_2r_3)^{1/3}$ turns out to be a convenient definition of the triangle scale, since it shows a tight correlation with the
measurement errors. Furthermore it separates well larger triangles for which the models perform well from smaller ones, for which the
measurements are presumably strongly affected  by higher order terms.
We found that the deviation between the perturbative model predictions of the different three-point correlations
from the measurements converge to the $1-2\sigma$ level for $(r_1r_2r_3)^{1/3} \gtrsim 60$ \mpc{}, while the
noisy error estimation imposes some uncertainty on this value. Note here that the smallest $r_i$ value above zero
in our analysis corresponds to the size of the $8$ \mpc{} grid cells into which we divided the simulations for computing
the correlations.
%
However, when the measurement errors are small (in particular their shot-noise contribution), which is the case for $Q_m$ and $Q_\times$,
and when the predictions are computed from the linear instead of the non-linear power spectrum, we find deviations above $2\sigma$
for $(r_1r_2r_3)^{1/3} \gtrsim 60$\mpc{}.
We attribute this effect to non-linearities around the BAO peak from large scale displacements and bias contributions not included in our treatment.
The fact that this effect is much weaker when using the de-wiggled, or the non-linear power spectrum indicates that the latter can
incorporate higher orders in the perturbative model for $Q$ to some degree.

Validating the model for the quadratic terms in $Q_h$ and $Q_\times$ with the $\Delta Q$ measurements, we found the predictions
based on the non-local bias model to show an overall better performance than those from the local model.
An exception are measurements in highly biased halo samples from small triangles with intermediate opening angles,
which are better described by the local than the non-local bias model. We interpret this effect as a compensation of
the non-local and the higher order terms not included in our bias treatment, which occurs for these particular triangles and
this particular halo sample.

Interestingly, the deviations of the $\Delta Q$ predictions based on the \bdsqfix{} bias model from the measurements are similar,
and for low biased samples even smaller than those based on the non-local model. For $(r_1r_2r_3)^{1/3} \gtrsim 60$\mpc{} the
significance of the deviations between $\Delta Q$ predictions and measurement is similar for all bias models, presumably
because of the low signal-to-noise ratio.

From the $Q_\times$ measurements we conclude that the leading order perturbative model predictions in combination with
the bias derived from the same statistics in Fourier space are a good approximation, with $1-2 \sigma$ deviations
$(r_1r_2r_3)^{1/3} \gtrsim 60$\mpc{}. These deviation are slightly higher around the aforementioned BAO feature,
but given the small errors on $Q_\times$ this agreement is still good. The model performance for $Q_h$ at large
scales is even better, despite the fact that terms beyond leading order, which are neglected
in the model should affect $Q_h$ more  strongly than $Q_\times$. This might be a result of the lower signal-to-noise ratio.
However, the $Q_h$ predictions differ significantly from the measurements for $(r_1r_2r_3)^{1/3} \lesssim 50$\mpc{}.
It is thereby important to note, that these results are specific to our small measurement errors from the combined
$\sim100 \ (h^{-1} \text{Gpc})^{3}$ of the $49$ simulations studied in this work. In practice, the deviation between model
predictions and measurements can be expected to be less significant, as the measurement errors are larger for the
smaller volumes of current and upcoming galaxy surveys.

As for $\Delta Q$, the $Q_h$ and $Q_\times$ predictions from the \bdsqfix{} model agree equally well with the measurements
at large scales for highly biased samples (high masses, high redshift). For low biased samples (low mass, low redshift)
this model describes the $Q_h$ and $Q_\times$ measurements even better than the non-local model. Differences in the
$Q_h$ predictions based on the linear and non-linear power spectrum are negligible compared to the larger measurement errors.

The good performance of predictions from the \bdsqfix{} model at larges scales suggests that a roughly universal $c_2(b_1)$ relation,
together with the local Lagrangian $g_2(b_1)$ relation, could tighten constraints on the linear bias, derived from third-order statistics in galaxy surveys.
However, recent studies pointed out, that assembly bias can lead to deviations from a universal $c_2(b_1)$ relation \citep{Modi16, Paranjape17}.
An application of the \bdsqfix{} model in the analysis of galaxy surveys therefore requires tests in mock catalogues
(for instance from semi-analytic models of galaxy formation) to validate for which type of galaxy samples these bias
relations are useful approximations.
More generally, the results presented in this paper show a good overall agreement of the non-local quadratic bias models
with simulations, using the same bias parameters for Fourier and configuration space, but the range of validity will
depend strongly on the samples used (volume, redshift and bias), so a detailed comparison with mock galaxy and corresponding
dark matter catalogues with redshift space distortions will be needed.

%% file: sections/appendix.tex
\section{3PCF binning} \label{app:3PCF_binning}

\begin{figure}
  \centering
   \includegraphics[width=60mm, angle = 270]{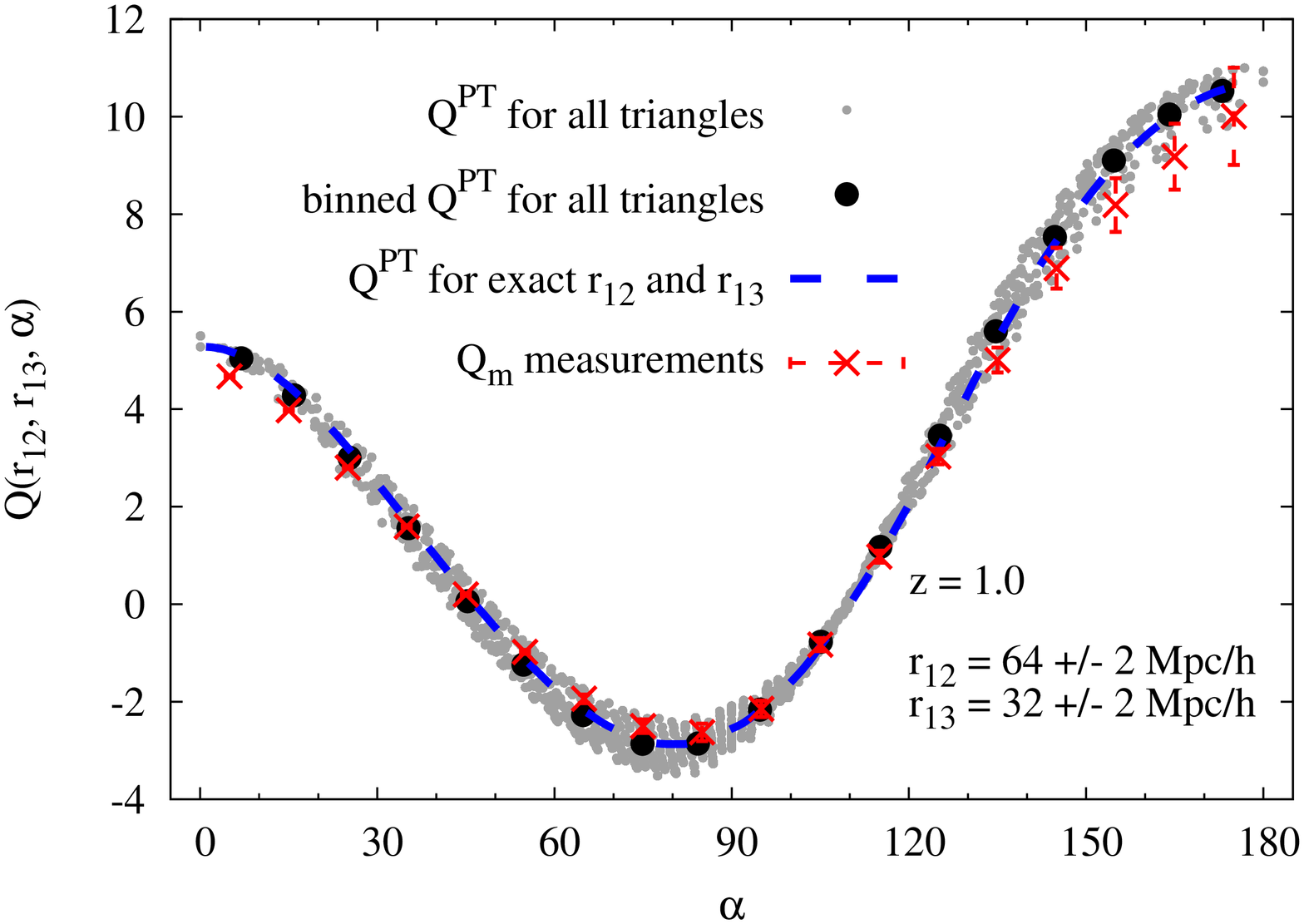}
   \caption{Testing the impact of the grid on the 3PCF. Grey dots show the predictions for the matter 3PCF,
   based on the measured power spectrum at $z=1.0$, for all triangles with $(r_{12}, r_{13}) \pm (\delta r_{12}, \delta r_{13})
   = (64, 32) \pm (2, 2)$ \mpc{} on the grid versus the opening angles $\alpha$. The mean predictions
   in bins of $\alpha$ are shown as black dots. Predictions for $(\delta r_{12}, \delta r_{13}) = (0,0)$
   (as we use them in our analysis) are shown as dashed blue line. Measurements of the 3PCF at $z=0.0$
   are shown with $1\sigma$ errors as red symbols.} 
\label{fig:3PCF_binning_6432}
\end{figure}

When measuring the 3PCF on a grid of cubical cells, we need
to allow for a tollerance of the triangle leg sizes $(r_{12}, r_{13})$ to
obtain a sufficiently larger number of triangles in each bin of the triangle
opening angle $\alpha$ (see Section \ref{subsec:3pc_sim}).
Here we test for different triangle configurations
how much the results are affected by this tolerance. We therefore compute the matter 3PCF prediction
for all triangles on the grid, which fulfill the condition $(r_{12}, r_{13}) \pm (\delta r_{12}, \delta r_{13})$.
As an example we show the results for the configuration $(64, 32) \pm (2, 2)$ \mpc{} in Fig. \ref{fig:3PCF_binning_6432}
as grey dots versus $\alpha$. The average 3PCF predictions in bins of $\alpha$ are shown as black
dots at the mean angle in each bin. These results are compared to predictions for exact values of $(r_{12}, r_{13})$
(i.e. $(\delta r_{12}, \delta r_{13}) = (0,0)$). We find that the difference between the two types of predictions
is small, compared to the difference between predictions and measurements (red symbols). We obtain the same results
for different triangle configurations (not shown here) and conclude that the binning of the 3PCF measurements
has no significant impact on the compariosn with the unbinned theory predictions, which we use in our analysis.


\section{Deviations for individual triangles \label{app:signif}}

\begin{figure}
  \centering
   \includegraphics[width=65mm, angle = 270]{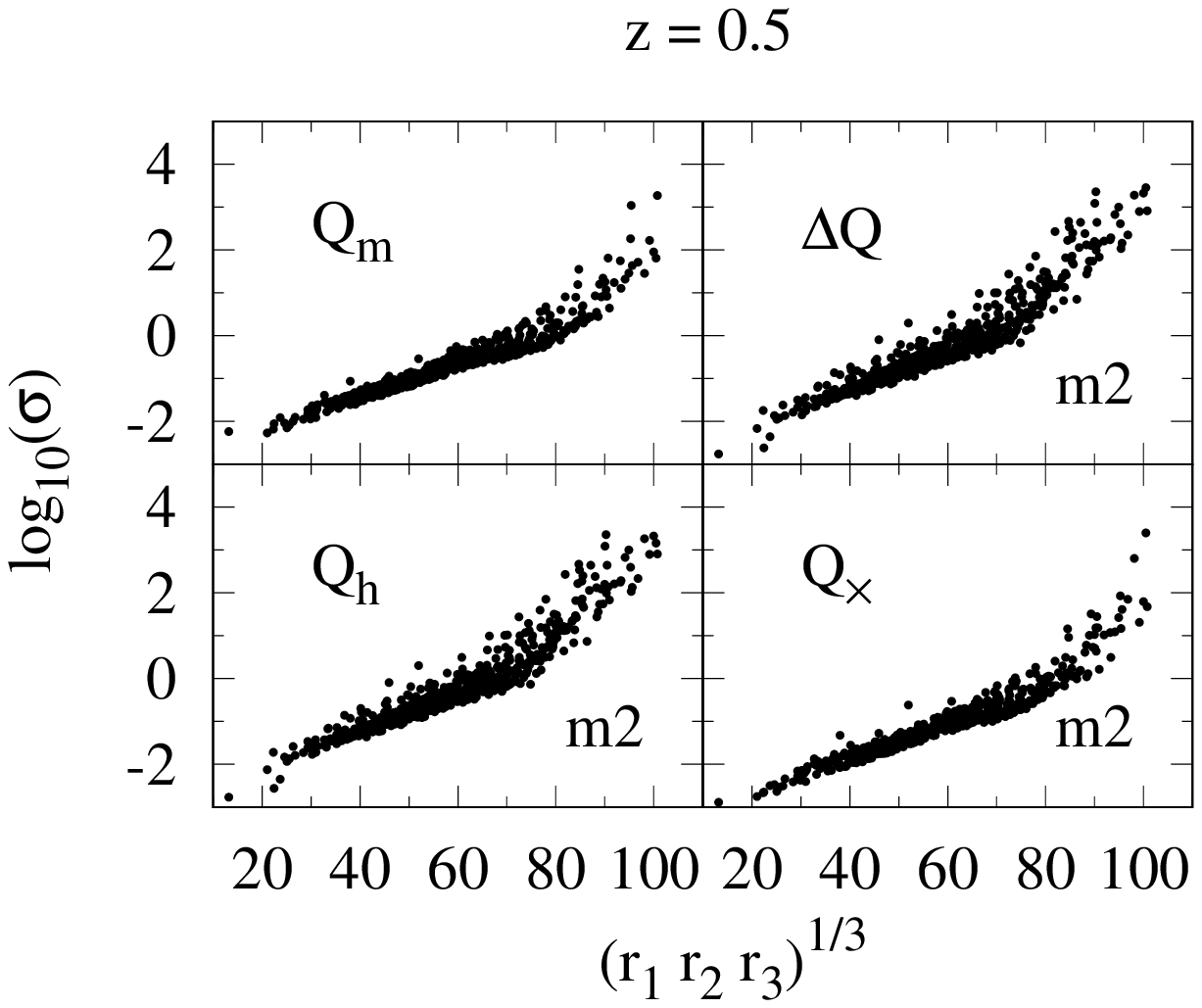}
   \caption{Examples of $1 \sigma$ errors of the different reduced 3PCFs studied in this paper versus the triangle scale. Results for different
   redshifts and mass samples are similar.}
\label{fig:Qm_error}
\end{figure}

The errors of the different $Q$ measurements correlates strongly with the total triangle scale, defined by $(r_1r_2r_3)^{1/3}$
as shown in Fig. \ref{fig:Qm_error}. We therefore study here the significance of the deviations between measurements
and predictions versus this scale.

\begin{figure}
  \centering
   \includegraphics[width=90mm, angle = 270]{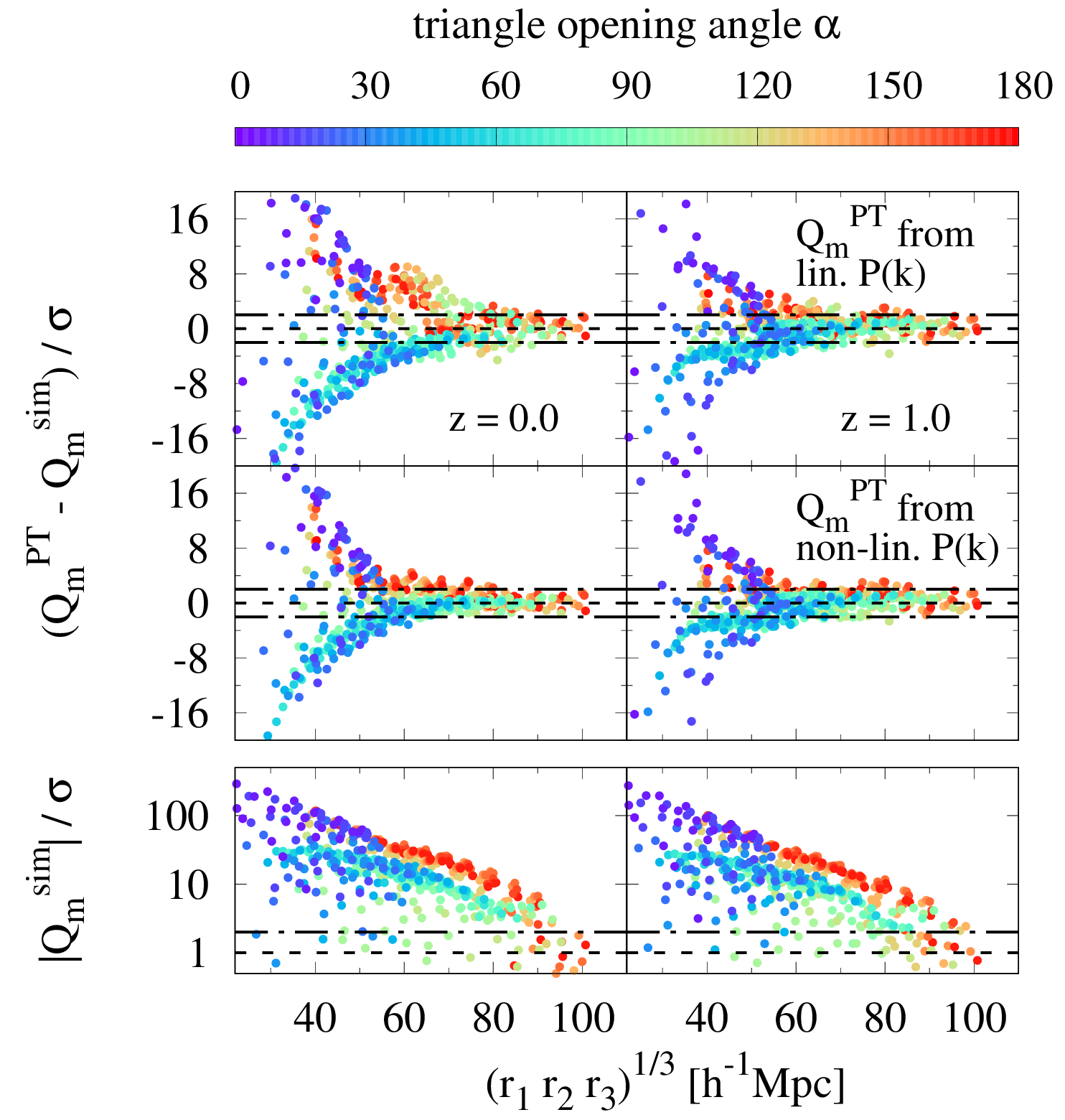}
   \caption{Top panels: Significance of the deviation between $Q_m$ measurements and tree-level predictions
   (based on the linear and non-linear power spectrum) versus the triangle scale.
   Dashed- dotted lines denote $2\sigma$ deviations. Bottom panel: Signal-to-noise ratio.
   Colours denote the triangle opening angle.}
   \label{fig:Qm_signifsimpt_r1r2r3}
\end{figure}

Fig. \ref{fig:Qm_signifsimpt_r1r2r3} shows that the $Q_m$ predictions
deviate from the measurements by less than $2 \sigma$ for
$(r_1r_2r_3)^{1/3} \gtrsim 80$ when predictions are computed from the linear power spectrum and $\gtrsim 60$ \mpc{}
when using the non-linear power spectrum. Note that these results are specific for the joint $\sim100 \ (\text{Gpc}/h)^3$
volume of the $49$ simulations. For smaller volumes (as covered by current galaxy surveys) errors would be larger and
the significance therefore smaller. Using alternative measures for the triangle scale, such as the triangle area or the sum
of the triangle legs leads to a less clear separation between triangles with weak and strong significance of the deviations.

\begin{figure}
  \centering
   \includegraphics[width=75mm, angle = 270]{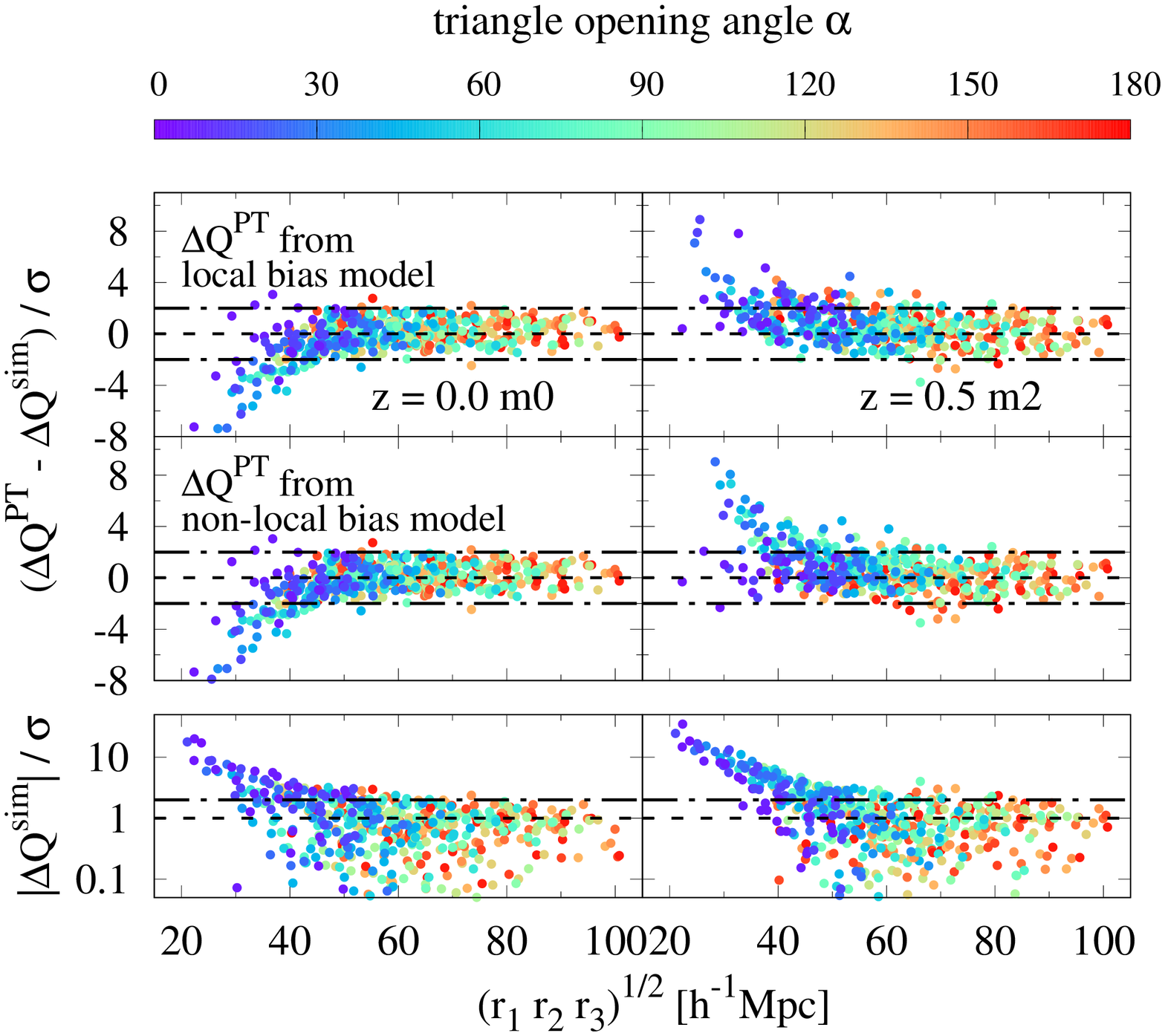}
   \caption{Significance of the deviation between predictions for $\Delta Q$
	and measurements versus triangle scale $(r_1r_2r_3)^{1/3}$
	for the halo mass samples m0 at $z=0.0$ and and m1 at $z=0.5$ (left and right panels respectively).
	The top and central panels show results for predictions from the local and non-local model respectively
	 (Table \ref{table:bias_models}), based on the non-linear power spectrum. The bottom panel shows
	 signal-to-noise ratio of measurements.}
\label{fig:dQ_signifsimpt_r1r2r3}
\end{figure}

The corresponding results for $\Delta Q$ are shown in Fig. \ref{fig:dQ_signifsimpt_r1r2r3}
for the low biased sample (m0) at $z=0.0$ and the highly biased sample (m2) at $z=0.5$
(with $b_1=1.43$ and $b_1=3.29$ respectively).
For the sample with the low linear bias the model predictions are below the measurements at \rtr{} $\lesssim 60$ \mpc{}.
Differences between local and non-local model predictions are not apparent, as expected from Fig. \ref{fig:dQ_signifsimpt_alpha_r1r2_proj}.
For the sample with the higher linear bias the predictions are above the measurements for \rtr{} $\gtrsim 60$ \mpc{} and the
non-local model performs slightly better than the local model at small scales. At large scales differences between model and
predictions are not signifiant for both samples, due to the low signal-to-noise ratio, which is shown in the bottom panel of
Fig. \ref{fig:dQ_signifsimpt_r1r2r3}. Note that the predictions are based on the non-linear power spectrum, measured in the simulation,
while the linear power spectrum leads to very similar results.

\begin{figure}
  \centering
   \includegraphics[width=75mm, angle = 270]{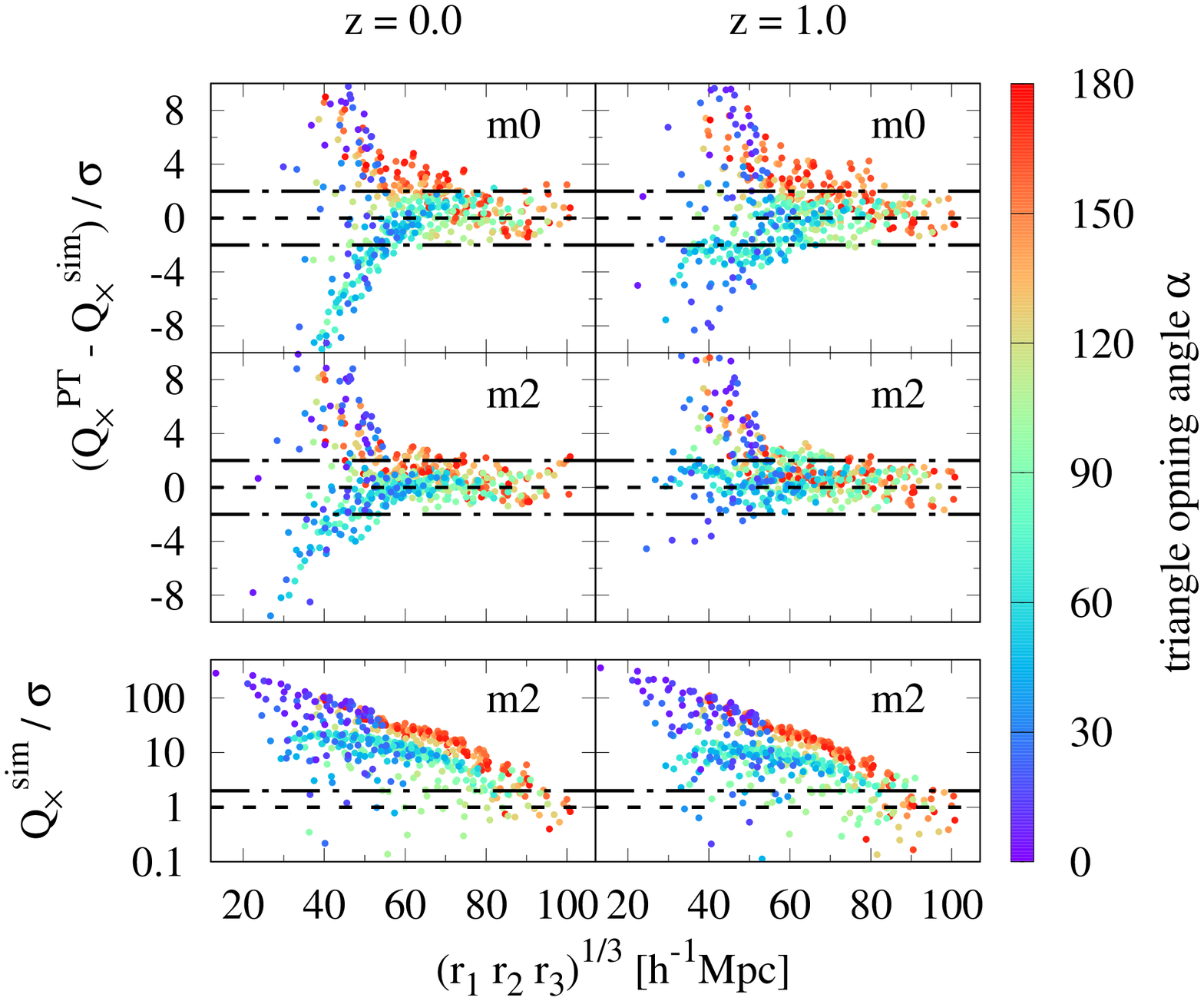}
   \caption{
   Significance of the deviations between predictions for $Q_\times$
	and measurements versus triangle scale $(r_1r_2r_3)^{1/3}$.
	The predictions are based on the non-local bias model (Table \ref{table:bias_models})
	and the non-linear power spectrum. Results are shown for the halo mass samples m0
	and m2 at $z=0.0$ and $z=1.0$ (left and right, top, bottom panels respectively).
	The lower sub panels show the signal-to-noise ratios
	 for the samples m2, which have higher shot-noise contributions than the m0 sample.  
   }
\label{fig:Qx_signifsimpt_r1r2r3}
\end{figure}

\begin{figure}
  \centering
   \includegraphics[width=75mm, angle = 270]{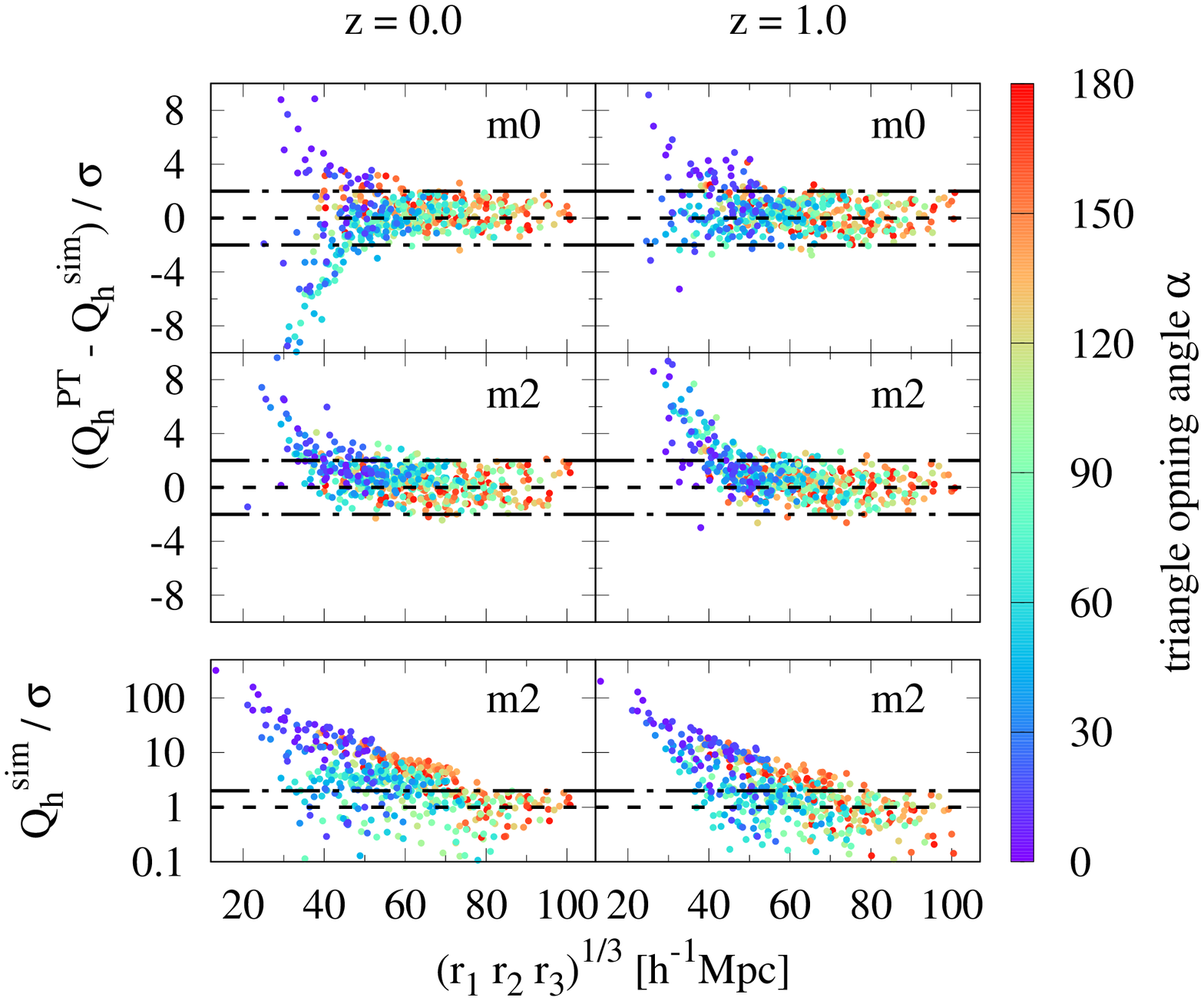}
   \caption{same as Fig. \ref{fig:Qx_signifsimpt_r1r2r3}, but for $Q_h$. 
   }
\label{fig:Qh_signifsimpt_r1r2r3}
\end{figure}

 The significance of the deviations between non-local bias model predictions for  $Q_h$ and $Q_\times$
 and the corresponding measurements are displayed versus the triangle scale in Fig. \ref{fig:Qx_signifsimpt_r1r2r3}
 and \ref{fig:Qh_signifsimpt_r1r2r3}.
 Covering a larger range of bias values ($1.43 \lesssim b_1 \lesssim 3.99$), we now show results for the mass samples m0 and m2, each at redshift
$z=0.0$ and $1.0$. Also here the predictions are based on the non-linear power spectrum and we find
very similar results when using the linear power spectrum.
The results are consistent with those shown in Fig. \ref{fig:Qhx_signifsimpt_alpha_r1r2_proj} as the predictions are most
significant for small triangles, where they show a strong dependence on the triangle opening angle for low biased samples,
while samples with high bias (higher masses and redshifts) show a weaker dependence on the opening angle
at small scales. Overall the deviations for both, $Q_\times$ and $Q_h$ converge to values of $\lesssim 2 \sigma$
for all samples for \rtr{}$\gtrsim 60$\mpc{}. An exception are results $Q_\times$ for large opening angles, which can
be attributed non-linearities around the BAO peak, as mentioned in the discussion of Fig. \ref{fig:Qhx_signifsimpt_alpha_r1r2_proj}.

\section{Covariances \label{app:cov}}

\begin{figure*}
  \centering
   \includegraphics[width=75mm, angle = 0]{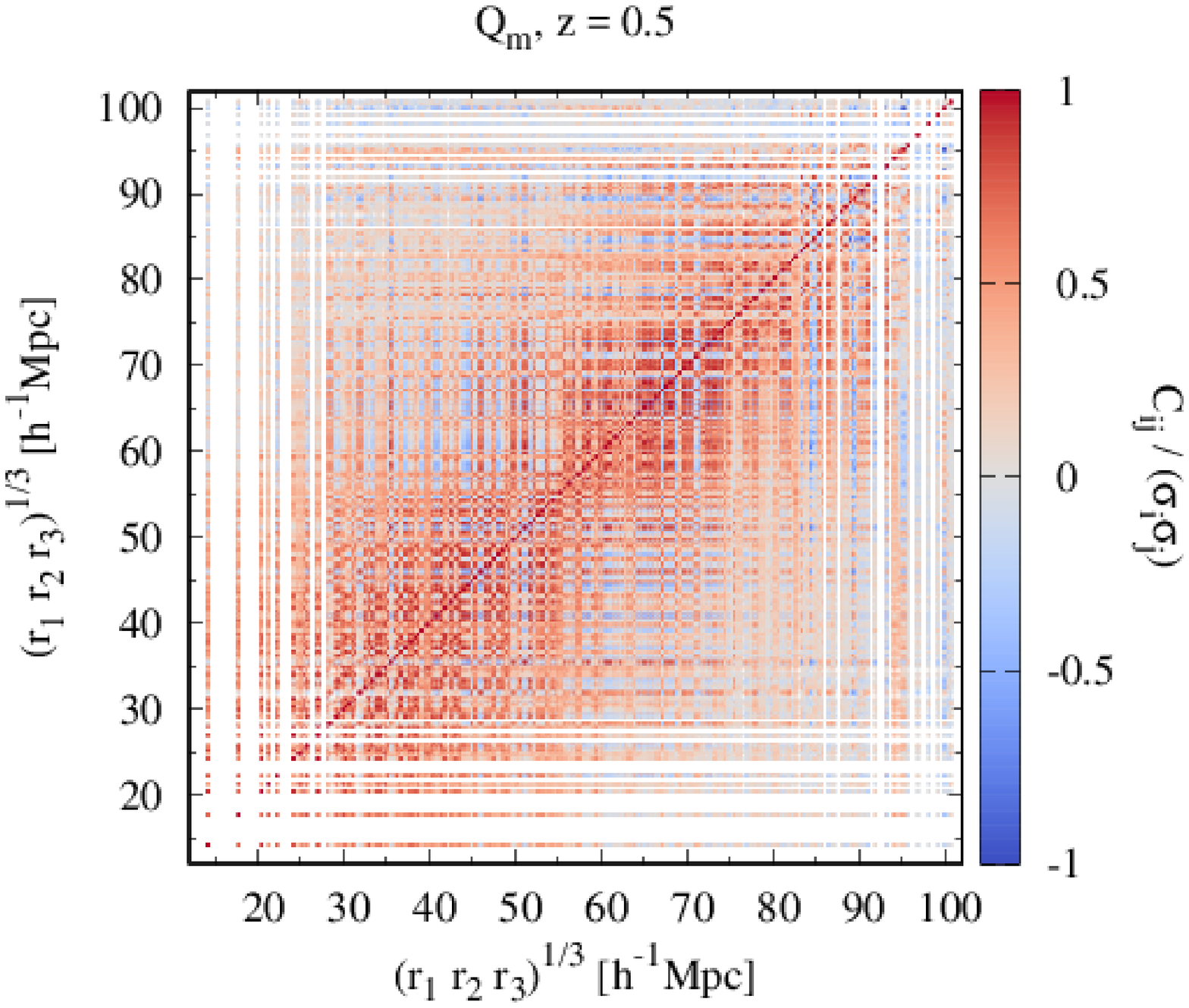}
   \includegraphics[width=75mm, angle = 0]{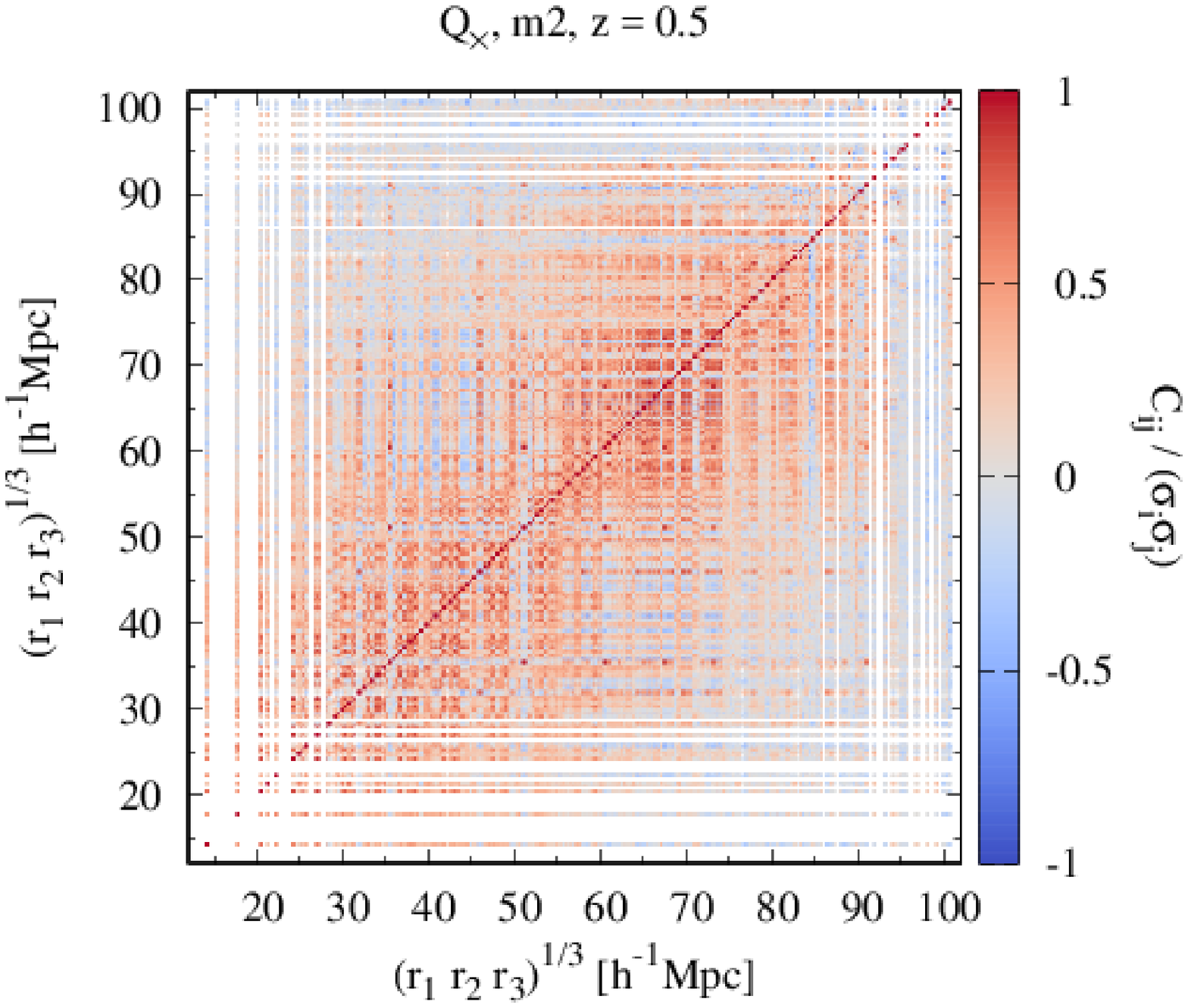}
   \includegraphics[width=75mm, angle = 0]{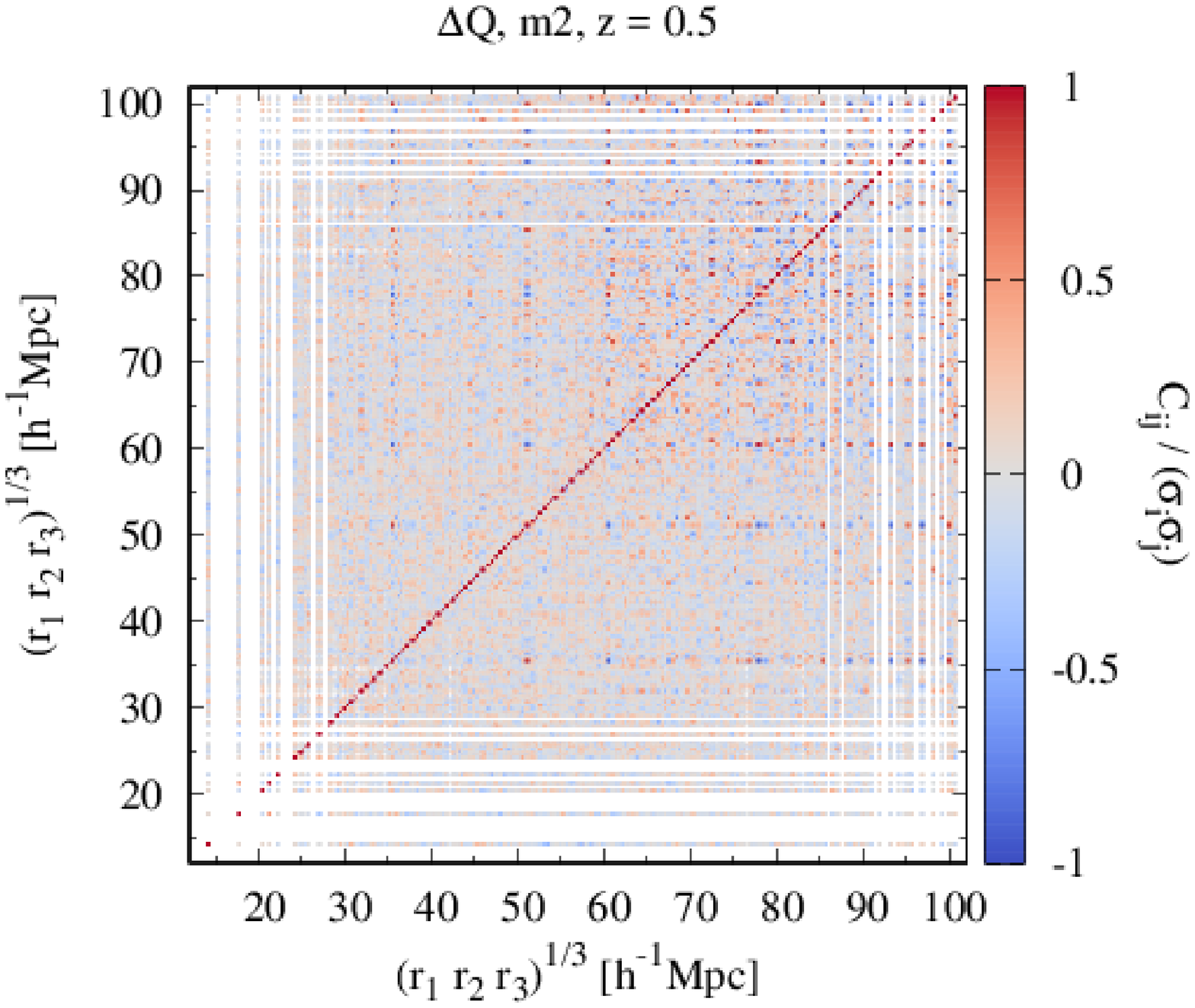}
   \includegraphics[width=75mm, angle = 0]{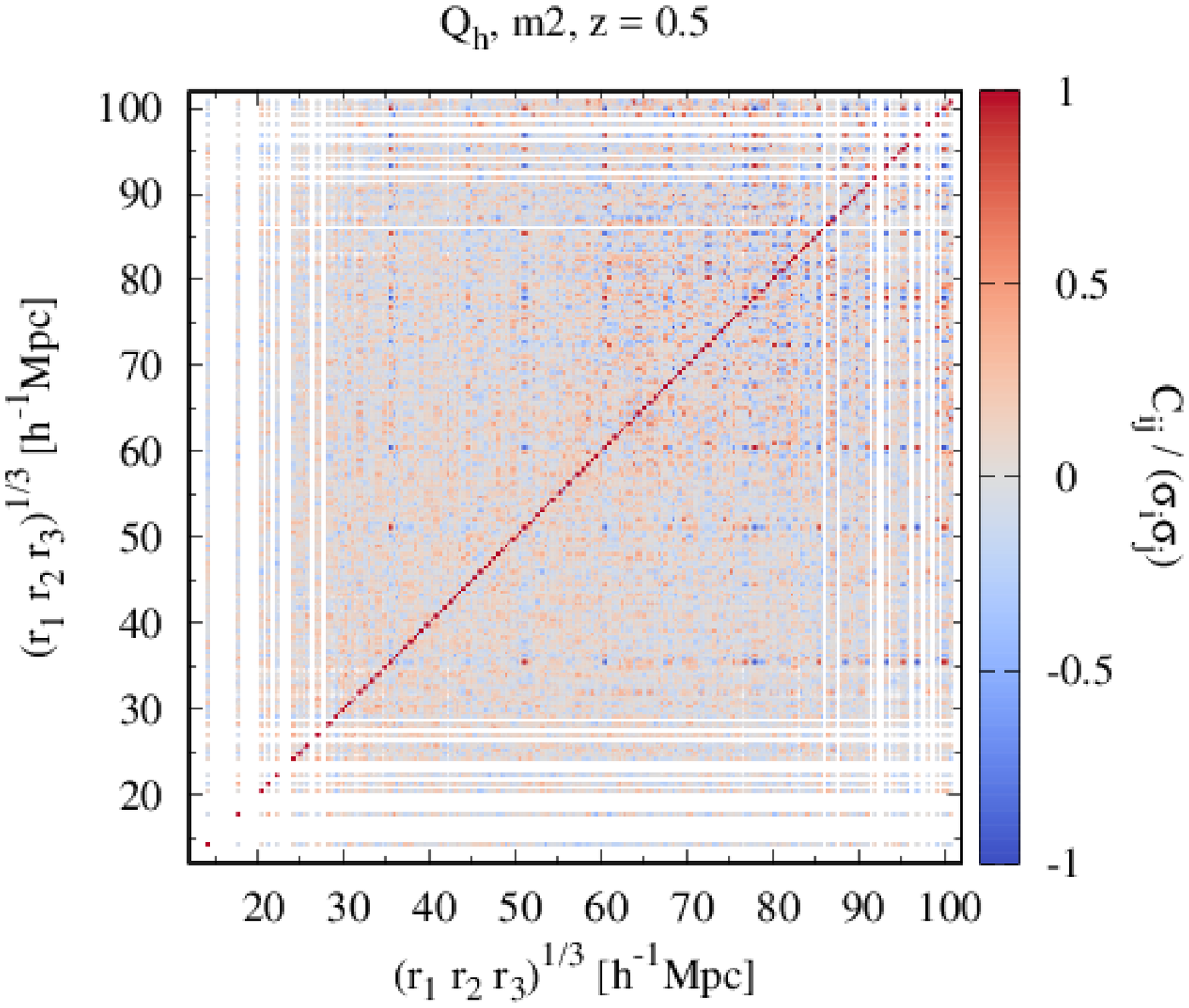}
   \caption{Examples of normalised covariances between different reduced 3PCS
   from the $504$ triangles used in this work. They were obtained from a set of $49$ simulations.
   The low amplitude of the off-diagonal elements in the  $Q_h$ covariances indicates
   a dominance of shot-noise errors.
   Results for $Q_h$ and $Q_\times$ are almost identical, because
   the $Q_\times$ errors are dominated by the $Q_h$ contribution.
   For computing the $\chi^2$ deviation from the model prediction we
    select triangles in scale bins containing $30$ triangles and perform a singular value
    decomposition, as described in Section \ref{subsec:3pc_sim}.}
\label{fig:Q_cov}
\end{figure*}

In Fig. \ref{fig:Q_cov} we show examples of the normalised covariance matrices at $z=0.5$ for the different three-point statistics
versus the triangle scale \rtr{}. The covariances for $Q_m$ and $Q_\times$ show strong off-diagonal elements,
while those of $Q_h$ and $\Delta Q$ are dominated by the diagonal elements, which indicates high shot-noise contributions.
Subsets of these covariances with $30^2$ elements around the diagonal are used for the $\chi^2$ estimation,
described in Section \ref{subsec:3pc_errorrs}.

\begin{figure*}
  \centering
   \includegraphics[width=60mm, angle = 270]{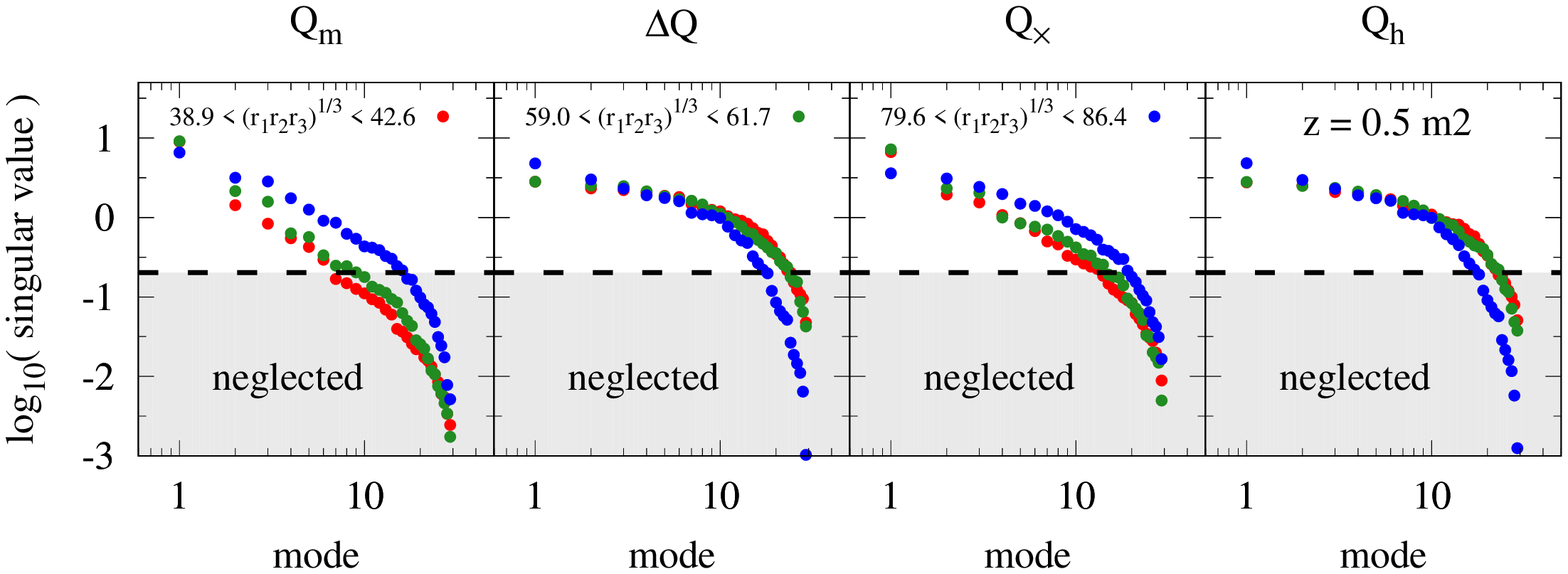}
   \caption{Singular values of the covariance matrices for the different three-point statistics studied in this work
   versus the mode number. The maximum mode number corresponds to the number of triangles
   in the \rtr{} bins. Results are shown for the mass sample m1 at $z=0.5$.
   Modes with singular values of less than $\lambda^2 \lesssim \sqrt{2/N_{sim}}$ are associated with noise
   and therefore neglected in the $\chi^2$ computation. The total number of modes is $30$, which corresponds
   to the number of triangles in each \rtr{} bin.}
\label{fig:covSV}
\end{figure*}

In order to reduce the impact of noise on these estimations we perform a singular value decomposition of the covariances.
The distribution of singular values is shown in Fig. \ref{fig:covSV} and reveals that a significant fraction of modes has
only a minor contribution to the covariance. One can see how the singular values for the shot-noise dominated covariances
of $\Delta Q$ and $Q_h$ show a slightly more pronounced drop, while those of the $Q_m$ and $Q_\times$ covariances
decay more slowly. We associate modes below $\lambda^2 \lesssim \sqrt{2/N_{sim}}$ with noise in the covariance
measurement and neglect them in the $\chi^2$ computation.

